\documentclass[fleqn,usenatbib]{mnras}
\usepackage{newtxtext,newtxmath}

\usepackage[T1]{fontenc}
\usepackage[all]{hypcap}

\DeclareRobustCommand{\VAN}[3]{#2}
\let\VANthebibliography\thebibliography
\def\thebibliography{\DeclareRobustCommand{\VAN}[3]{##3}\VANthebibliography}

\usepackage{graphicx}   
\usepackage{amsmath}    
\usepackage{enumitem} 
\usepackage{multirow}

\title[Surviving the Heat]{Surviving the Heat: multi-wavelength analysis of V883 Ori reveals that dust aggregates survive the sublimation of their ice mantles}

\author[Houge, Mac{\'\i}as \& Krijt]{
Adrien Houge$^{1, 2}$\thanks{E-mail: ah1006@exeter.ac.uk}, Enrique Mac{\'\i}as$^{2}$, and Sebastiaan Krijt$^{1}$
\\
$^{1}$Department of Physics and Astronomy, University of Exeter, Exeter, EX4 4QL, UK\\
$^{2}$European Southern Observatory, Karl-Schwarzschild-Strasse 2, D-85748 Garching bei Munchen, Germany}

\date{Accepted 2023 December 4. Received 2023 December 1; in original form 2023 August 3}

\pubyear{2023}

\begin{document}
\label{firstpage}
\pagerange{\pageref{firstpage}--\pageref{lastpage}}
\maketitle

\begin{abstract}

Investigating the response of icy dust aggregates to water ice sublimation is essential for understanding the formation and properties of planetesimals in protoplanetary discs. However, their fate remains unclear, as previous studies suggest aggregates could either survive or {completely fall apart} to (sub)$\mathrm{\mu m}$-sized grains. Protoplanetary discs around stars undergoing accretion outbursts represent a unique laboratory to study the ice sublimation process, as the water snowline is pushed outward to regions accessible to current observatories. In this work, we aim to understand the aggregates' response to ice sublimation by focusing on V883 Ori, a system currently undergoing a powerful accretion outburst. We present new analysis of archival high resolution ALMA observations of the protoplanetary disc of V883 Ori at 0.88, 1.3, 2.0, and 3.1 mm, and derive new radial spectral index profiles, which we compare with predictions from one-dimensional dust evolution simulations. In the region of V883 Ori where water ice has sublimated, we find lower spectral indices than previously obtained, indicating the presence of cm-sized particles. Coupled with our dust evolution models, we find that the only way to explain their presence is to assume they formed before the outburst, and survived the sublimation process. The resilience of dust aggregates to such intense events leads us to speculate that it may extend to other environments with more gentle heating, such as pebbles drifting through the water snowline in quiescent protoplanetary discs. In that case, it may alter the formation pathway of dry planetesimals interior to the snowline.

\end{abstract}




\section{Introduction}
\label{sec:introduction}

Planets form in discs of dust and gas surrounding young stars. Inherited from the collapsed molecular cloud, the initial reservoir of solids is made of $\mu$m-sized dust grains, which coagulate into larger and larger aggregates through sticky collisions at low velocities \citep[]{weidenschilling1993formation}. When particles reach $\approx$ cm-size, so-called pebbles, the growth is halted for example due to fragmenting collisions or efficient inward radial drift \citep[]{blum1993experimental, weidenschilling1997origin, blum2008growth}. {The streaming instability (SI) is a promising mechanism to overcome the growth barrier and achieve the formation of km-sized planetesimals, by forming highly concentrated clumps of pebbles that become gravitationally unstable \citep[]{youdin2005streaming, johansen2007rapid, johansen2009particle, bai2010dynamics}. But to be triggered, the SI requires regions of protoplanetary discs with a pre-existing enhanced dust-to-gas ratio, as compared to the typical value of $1$ per cent. 
\citep[]{johansen2009particle, bai2010dynamics, carrera2015form, li2021thresholds}.

In that context, the water snowline is thought to play an important role, as planetesimals have been argued to form preferentially (a) interior to the snowline, because of the reduced stickiness of dry silicates as compared to ice-rich grains which leads to a traffic jam effect \citep[]{dominik1997physics, supulver1997sticking, wada2013growth, gundlach2014stickiness}, and (b) just exterior to the snowline because of the cold finger effect increasing the ice surface density \citep[]{stevenson1988rapid, cuzzi2004material, ros2013ice, drkazkowska2017planetesimal, schoonenberg2017planetesimal, lichtenberg2021bifurcation}. Formation route (a) has been recently called into question, with the traditional view of a lower $v_\mathrm{frag}$ for silicates as compared to ice-rich particles being challenged by new laboratory experiments \citep[e.g.][]{gundlach2018tensile, musiolik2019contacts}. If, however, icy pebbles happen to disintegrate into $\mu$m-sized grains upon drifting through the snowline \citep[many-seeds scenario, see e.g.][]{schoonenberg2017planetesimal}, their drift velocity would still be reduced, offering another pathway for the traffic jam effect and the formation of dry planetesimals. Interestingly, the release of $\mu$m-sized grains would also boost formation route (b), as small particles can diffuse back across snowline and enhance further the solids-to-gas ratio \citep[]{saito2011planetesimal, ida2016radial, schoonenberg2017planetesimal, hyodo2019formation, hyodo2021planetesimal}. In summary, the response of icy dust aggregates to sublimation plays an important role in dust evolution and planetesimal formation around the snowline. However, the water snowline is generally located relatively close to its host star \citep[e.g.][]{mulders2015snow}, in regions inaccessible to current observatories at millimetre wavelengths, which prevents us to probe the state of the dust size distribution.

A unique opportunity to study how the water snowline and ice sublimation affect the dust population is offered by protoplanetary discs undergoing FUor-type accretion outbursts. During such events, the luminosity and mass accretion rate of the central protostar increase by ${\sim}2{-}3$ orders of magnitude, and can remain elevated for hundreds of years \citep[]{kenyon1990iras, dunham2012resolving, audard2014episodic, fischer2022accretion}. As a result, the dust and gas content in the surrounding protoplanetary disc significantly warms up, leading for example to the crystallization of amorphous silicates in the inner disc \citep[]{abraham2009episodic}, and sublimation of ice species on large scales as snowlines effectively move outwards \citep[e.g.][]{jorgensen2013recent, molyarova2018chemical, wiebe2019luminosity}.

A few protoplanetary discs are known to be currently undergoing FUor-type accretion outbursts \citep[]{fischer2022accretion}. Notably, V883 Ori is a young $1.3~\mathrm{M_\odot}$ star surrounded by a $0.2 - 0.6~\mathrm{M_\odot}$ protoplanetary disc still embedded in its envelope \citep[]{cieza2016imaging, cieza2018alma, leemker2021chemically}, undergoing a powerful FUor-type accretion outburst that likely began before 1888 \citep[]{pickering1890detection}. The system received a lot of attention with ALMA in the last few years, including a suite of spectral line observations \citep[e.g.][]{van2018methanol, lee2019ice, leemker2021chemically, ruiz2022chemical, tobin2023deuterium}. A first analysis of the water snowline of V883 Ori was made by \citet[]{cieza2016imaging}, using dust continuum emissions exclusively in ALMA Band 6. The continuum data revealed a dark annulus at $42 \mathrm{~au}$ as well as an abrupt change in optical depth and millimetre spectral index. These features matched prediction of steady state (i.e. non-outbursting) dust and snowline models \citep[]{banzatti2015direct}, prompting Cieza et al. to conclude the snowline in V883 Ori had been moved significantly outwards by the outburst. More recently, \citet[]{tobin2023deuterium} estimated the water snowline to be located at $80 \mathrm{~au}$ based on direct measurements of the column density of HDO and $\mathrm{{H_2} ^{18}O}$.

Studies of the dust size distribution in discs undergoing accretion outbursts suggests different scenarios for the response of icy aggregates to the sublimation of their ice content. Based on ALMA Band 6 observations of V883 Ori presented in \citet[]{cieza2016imaging}, \citet[]{schoonenberg2017pebbles} developed a monodisperse dust coagulation and sublimation model and obtained close agreement to the measured spectral index with a model in which: (1) icy aggregates are completely disrupted following sublimation at the outburst onset, and (2) their $\mathrm{\mu m}$ silicate seeds rapidly re-coagulate to the local fragmentation threshold of $300 \mathrm{~\mu m}$. Conversely, \citet[]{liu2021millimeter} employed multi-wavelength observations to infer the presence of large pebbles ($\gtrsim 1.6 \mathrm{~mm}$) in the inner $\sim 10 \mathrm{~au}$, $400 \mathrm{~K}$ hot disc around FU Ori, also currently undergoing an accretion outburst, indicating that here pebbles had survived the sublimation process. An alternative way of studying the question of the dust response to ice sublimation comes from laboratory experiments, in which mixtures of dust/ice can be heated and studied in controlled settings. Previous laboratory studies also found a variety of outcomes, including complete disintegration of the dust aggregates back to $\mu$m silicate seeds \citep[]{aumatell2011breaking}, or survival of the refractory matrix \citep[]{spadaccia2022fate}.

In this paper, we aim to shed some light on how icy dust aggregates respond to water ice sublimation by performing a new multi-wavelength analysis of V883 Ori, using archival ALMA data (see Sect.~\ref{sec:observations}) and the latest insights concerning the location of the water snowline \citep[]{tobin2023deuterium}. We select archival ALMA data at $0.88 \mathrm{~mm}$ (Band 7), $1.3 \mathrm{~mm}$ (Band 6), $2.0 \mathrm{~mm}$ (Band 4) and $3.1 \mathrm{~mm}$ (Band 3). Given that the snowline is generally located at $\approx 3 \mathrm{~au}$ in quiescent conditions, we focus on characterising the dust size distribution of V883 Ori in between $3 < r < 80 \mathrm{~au}$, where the dust population underwent sublimation at the onset of the outburst. From the multi-wavelength dataset, we robustly retrieve the radial profile of the dust emission spectral index (see Sect.~\ref{sec:observational_results}), a key tracer for dust particle size\footnote{In this manuscript, dust of any size will be referred to as particles, solids, or aggregates. We will use ‘grains’ when specifically targeting small solids (i.e. ${<}10 \mathrm{~\mu m}$), and ’pebbles’ for large solids (i.e. ${>}1 \mathrm{~mm}$).}. We additionally perform 1D dust coagulation and evolution simulations using the state-of-the-art \texttt{DustPy} package \citep[]{stammler2022dustpy}, adapted to include the effects of an accretion outburst. Given the variety of outcomes uncovered by laboratory studies \citep[]{aumatell2011breaking, spadaccia2022fate}, we incorporate two end-member scenarios for the response of icy particles to water ice sublimation: the \emph{many-seeds scenario}, in which all particles fall apart to $\mathrm{\mu m}$-sized grains after sublimation; and the \emph{resilient scenario}, in which dust particles survive (see Sect.~\ref{sec:simulation_methods}). In Section \ref{sec:dustpy_spectralindex}, we present the spectral index radial profiles from the dust evolution models in the two scenarios, and compare them qualitatively with the observational results. We discuss our main findings and their implications in Section \ref{sec:discussion}, followed by our conclusions in Section \ref{sec:conclusions}.

\section{Observations and data reduction}
\label{sec:observations}
\label{sec:selfcalibration}

To characterise the dust size distribution in V883 Ori, it is crucial to perform a multi-wavelength analysis with high resolution data. We gathered from the ALMA archive the highest resolution observations available in Band 7 (project code: 2016.1.00728.S, PI: Lucas Cieza), Band 6 (project code: 2015.1.00350.S, PI: Lucas Cieza), Band 4 (project code: 2017.1.00388.S, PI: Hauyu Baobab Liu), and Band 3 (project code: 2016.1.00728.S, PI: Lucas Cieza). In particular, the Band 6 data correspond to those previously used by \citet[][]{cieza2016imaging} to obtain the first estimate of the spectral index in V883 Ori, and discover the dark annulus at $42 \mathrm{~au}$. A detailed analysis of the Band 7 and 3 data will be presented in Cieza et al. (in prep.). As the high resolution data in Band 7 and Band 6 did not well cover the shorter baselines, we additionally retrieved lower resolution archival data in these two bands (resp. project code: 2015.1.00041.S, PI: John Tobin, and project code: 2013.1.00710.S, PI: Lucas Cieza). Additional details about the observations can be found in Table \ref{tab:obs_dataset}. {We note that while the observing dates differ by up to $2$ years, we do not expect the thermal dust emission of FU Orionis objects like V883 Ori to vary on such short timescales \citep[e.g.][]{wendeborn2020study}.}

The data were initially calibrated by ALMA staff, some manually and some using the ALMA pipeline. We used \texttt{CASA} \citep[Common Astronomy Software Applications v6.4.1;][]{mcmullin2007casa} for further inspection, self-calibration, and imaging. As a first step, we aligned the data by fitting a Gaussian to the image of each individual execution and shifting the phase center of each measurement set. Mismatches in the amplitude scales of different observations may also appear. We therefore rescaled the fluxes of each band following the methods in \citet[]{andrews2018disk}, considering one of the executions at each band as reference and scaling the others by a factor using the task \texttt{gencal}. 
The Band 7 observations have slightly different spectral setups between the long and short baseline data (see Table~\ref{tab:obs_dataset}), so in this band we only scaled the fluxes between the three short baseline executions.

We then performed self-calibration on each individual dataset; except on the short baselines of Band 7, where we started from the concatenation of the three executions. For the self-calibration, we employed the task \texttt{tclean} with the \texttt{hogbom} deconvolver, a \texttt{briggs} weighting with $robust=0.5$, and $nterms=1$, i.e. assuming a linear dependency for the emission frequency. We started with phase self-calibration following an iterative approach, decreasing step by step the solution intervals until it no longer improved the signal-to-noise ratio (S/N). This was followed by a round of amplitude self-calibration ($calmode=ap$), which did not provide any improvements of the S/N and was therefore not considered. Overall, self-calibration on individual datasets provided improvements of the S/N ranging in between $10$ and $220\%$. We then combined the short and long baseline executions of Band 7 and 6.

After combining the short and long baseline executions, we performed an additional cycle of phase self-calibration to ensure a good alignment of the data. This was followed by a round of amplitude self-calibration, especially important for the Band 7 dataset whose fluxes were not rescaled due to the frequency offset in the spectral windows. We used the task \texttt{tclean} now using $nterms = 2$, i.e. assuming the frequency dependency follows a Taylor expansion to quadratic terms. This second round of self-calibration provided an improvement of the S/N in Band 7 and 6 respectively by $13$ and $32\%$.

We then built the final images with the task \texttt{tclean}, using the \texttt{mtmfs} deconvolver {\citep[]{rau2011multi}}, $nterms = 2$, multiple scales from the point-source to $5$ times the beam size, and a \texttt{Briggs} weighting with different $robust$ values to explore higher resolutions. In addition {to the continuum images, we derived the dust emission spectral index $\alpha = \partial \ln{I_\nu}/\partial \ln{\nu}$, which is a widely used metric to trace the size distribution of dust particles in protoplanetary discs \citep[e.g. see reviews by][]{ testi2014dust, andrews2020observations}. To do so, we built spectral index maps by combining pairs of bands in the task \texttt{tclean}, which derives the spectral index at each pixel based on the observed visibilities, again using the \texttt{mtmfs} deconvolver and $nterms = 2$. This allowed us to obtain three spectral index maps from the four bands, while also benefiting from a better $uv$ coverage as compared to deriving the spectral index from the intensity measured in individual bands \citep[see also][]{huang2018co}. More details about the final images can be found in Table \ref{tab:final_images_info}.}

\begin{table*}
    \centering
    \caption{Summary of the ALMA observations.}
    \label{tab:obs_dataset}
    \begin{tabular}{cccccc}
        \hline \hline Project code & PI & {Observation} date & On-source time (min) & Baseline (m) & Frequencies (GHz) \\
        \hline
        Band 7  \\
        2016.1.00728.S & Lucas Cieza & 08 Sep. 2017 & $16.7$ & $21-10\,635$ & $333.4, 335.4, 345.4, 347.4$\\ 
        2015.1.00041.S & John Tobin & 06 Sep. 2016 & $0.3$ & $15 - 2\,483$ & $332.9, 343.9, 345.7$\\
         &  & 06 Sep. 2016 & $0.3$ & $16 - 2\,732$ & $332.9, 343.9, 345.7$\\
         &  & 19 Jul. 2017 & $0.3$ & $18 - 3\,696$ & $332.9, 343.9, 345.7$\\
        \hline
        Band 6 \\
        2015.1.00350.S & Lucas Cieza & 27 Oct. 2015 & $22.5$ & $84 - 16\,196$ & $218.0, 232.6$ \\
        2013.1.00710.S & Lucas Cieza & 30 Aug. 2015 & $3.2$ & $15 - 1\,466$ & $218.0, 232.6$\\
        \hline
        Band 4 \\
        2017.1.00388.S & Hauyu Baobab Liu & 07 Nov. 2017 & $41.7$& $113 - 13\,894$ & $146.0, 147.8, 158.2, 160.0$\\
        \hline
        Band 3 \\
        2016.1.00728.S & Lucas Cieza & 21 Sep. 2017 & $17.0$ & $41 - 12\,145$ & $90.5, 92.4, 102.5, 104.5$\\
        \hline
    \end{tabular}
\end{table*}

\begin{table*}
    \centering
    \caption{Final images and spectral index maps.}
    \label{tab:final_images_info}
    \begin{tabular}{cccccc}
        \hline \hline Band & Central wavelength (mm) & Robust {parameter} & RMS ($\mu \mathrm{Jy / beam}$) & Peak S/N & Beam shape \\
        \hline
        7 & $0.88$ & $0$ & $19$ & $374$ & $0.\!\!^{\prime\prime}041 \times 0.\!\!^{\prime\prime}019$, $\mathrm{PA} = 58^\circ$ \\
        6 & $1.33$ & $0$ & $26$ & $412$ &$0.\!\!^{\prime\prime}044 \times 0.\!\!^{\prime\prime}038$, $\mathrm{PA} = 77^\circ$ \\
        4 & $1.96$ & $0.5$ & $41$ & $331$ & $0.\!\!^{\prime\prime}047 \times 0.\!\!^{\prime\prime}030$, $\mathrm{PA} = 55^\circ$ \\
        3 & $3.07$ & $0.5$ & $61$ & $399$ & $0.\!\!^{\prime\prime}082 \times 0.\!\!^{\prime\prime}073$, $\mathrm{PA} = 37^\circ$ \\

        \hline
        Spectral index maps & & & \\
        7 + 6 & $1.06$  & $0$ & $\times$ & $\times$ & $0.\!\!^{\prime\prime}040 \times 0.\!\!^{\prime\prime}023$, $\mathrm{PA} = 56^\circ$ \\
        6 + 4 & $1.58$  & $0$ & $\times$ & $\times$ & $0.\!\!^{\prime\prime}044 \times 0.\!\!^{\prime\prime}037$, $\mathrm{PA} = 61^\circ$ \\
        4 + 3 & $2.39$  & $0.5$ & $\times$ & $\times$ & $0.\!\!^{\prime\prime}064 \times 0.\!\!^{\prime\prime}059$, $\mathrm{PA} = 49^\circ$ \\
        \hline
    \end{tabular}
\end{table*}

\section{Observational results}
\label{sec:observational_results}

\subsection{Final ALMA images}
\label{sec:results_final_images}
We present in Fig.~\ref{fig:fig1_final_images} our final images of the dust thermal emission of the protoplanetary disc around V883 Ori at $0.88 \mathrm{~mm}$ (Band 7), $1.3 \mathrm{~mm}$ (Band 6), $2.0 \mathrm{~mm}$ (Band 4) and $3.1 \mathrm{~mm}$ (Band 3). In Band 6, our image is obtained from the same dataset presented in \citet[]{cieza2016imaging}, although here combined with another execution to better cover the shorter baselines. In all bands, the dust emission appears azimuthally symmetric. A ring-shaped structure appearing as a dark annulus is well visible at $42 \mathrm{~au}$ in Band 7 and Band 6. This feature was also seen in Band 6 by \citet[]{cieza2016imaging}. The dark annulus is also slightly visible in Band 4, but absent in Band 3, which we explain by the lower resolution of this observation (Table \ref{tab:final_images_info}).

\begin{figure*}
    \includegraphics[width=\textwidth]{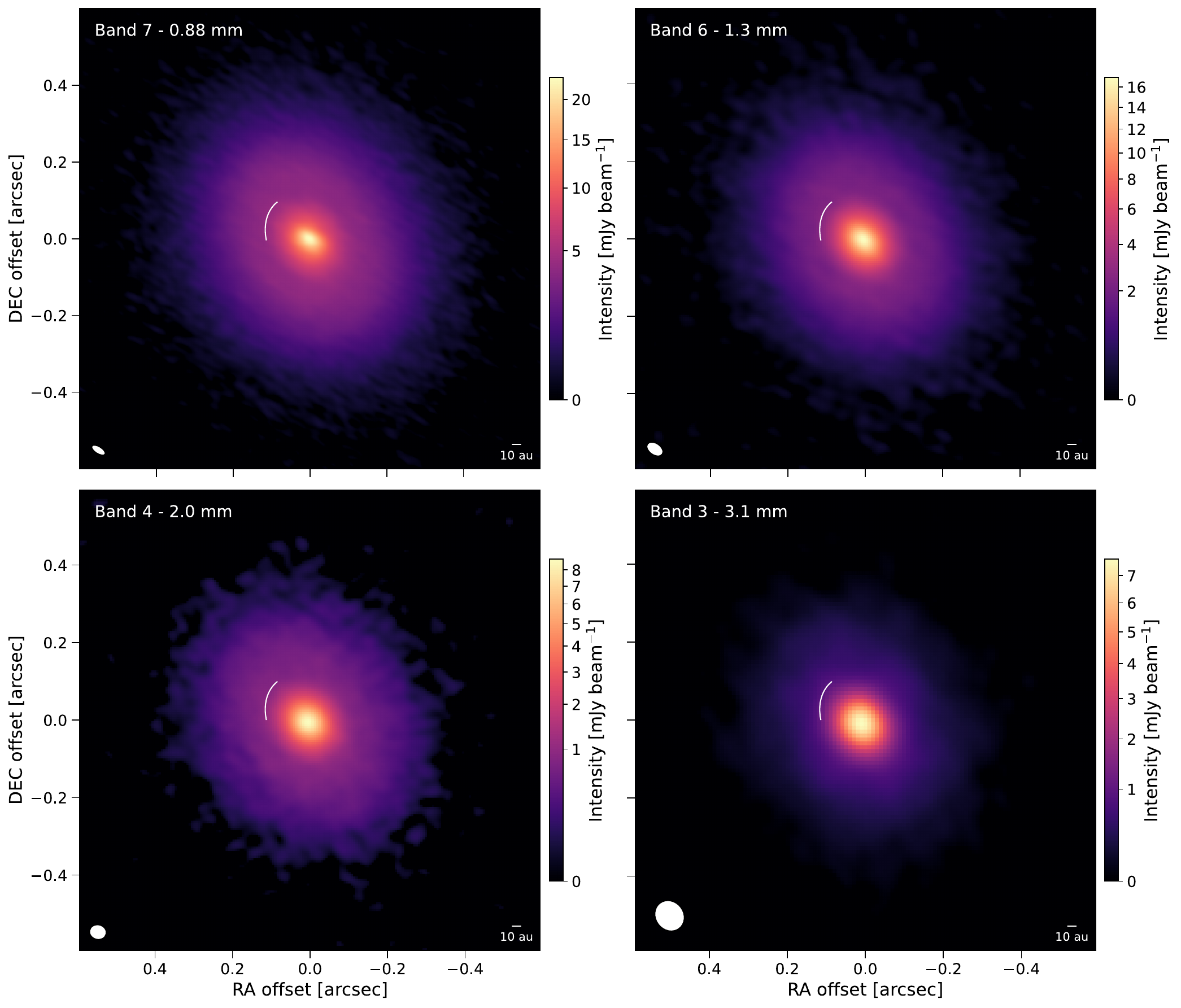}
    \caption{Final images of the dust thermal emission of the disc around V883 Ori at $0.88$ (top-left), $1.3$ (top-right), $2.0$ (bottom-left) and $3.1 \mathrm{~mm}$ ({bottom-right}). {The white curve indicates the dark annulus at $42 \mathrm{~au}$.} The color bar in each panel goes from about twice the rms up to the maximum intensity. More information about the final $robust$ values used, the rms, peak S/N, and beam shape can be found in Table \ref{tab:final_images_info}.}
    \label{fig:fig1_final_images}
\end{figure*}

\subsection{Spectral index radial profiles}
\label{sec:obs_results_spectral_index}

We present in Fig.~\ref{fig:fig3_SpInd_profile} the radial profiles of the dust {emission} spectral index at $0.88-1.3 \mathrm{~mm}$ (red curve), $1.3-2.0 \mathrm{~mm}$ (orange curve), and $2.0-3.1 \mathrm{~mm}$ (green curve). They are obtained from the spectral index maps mentioned in Sect.~\ref{sec:selfcalibration}, where we average the spectral index in elliptical rings of width $5 \mathrm{~mas}$ with an eccentricity given by the disc inclination $i=38.3^\circ$ and position angle $\mathrm{P.A.} = 32.4^\circ$ \citep[]{cieza2016imaging}. At each radius, the spectral index is given by the mean $\bar{\alpha}$ in the ring, and the uncertainty is calculated as the error of the mean following \citep[]{macias2021characterizing}
\begin{equation}
    \sigma_\mathrm{\bar{\alpha}} = \dfrac{\sigma_i}{\sqrt{A_i/A_\mathrm{beam}}},
\end{equation}

where $\sigma_i$ is the standard deviation in the ring, $A_i$ is the area of the ring, and $A_\mathrm{beam}$ is the area of the beam. We additionally show the azimuthal average of the spectral index error maps produced by \texttt{tclean} in fainter shades. In the lower right of each panel, we illustrate as solid grey lines the absolute flux calibration systematic errors of ALMA, assuming the nominal errors of $10\%$ in Band 7, and $5\%$ in Band 6, 4, and 3. Such systematic uncertainty could increase or decrease the spectral index at all radii up to a maximum shift indicated by the grey lines, but it would not affect the relative shape of the individual profiles. Our multi-wavelength dataset also suggests that these systematic errors should not be significant, as an important error in one band would lead to {shifts} in the other profiles, which is not visible in Fig.~\ref{fig:fig3_SpInd_profile}. 

The three spectral index profiles display common features. The inner disc ($r < 20 \mathrm{~au}$) is characterised by a region with $\alpha \lesssim 2$, distinctive of optically thick emissions with high dust albedo, in which case the emission is fainter than the black-body radiation \citep[e.g.][]{birnstiel2018disk}. It is followed by a transition ($20 < r < 50 \mathrm{~au}$) into the optically thin outer disc ($r > 50 \mathrm{~au}$) characterised by $\alpha \approx 2.5$. In this region, we notice that the dark annulus visible in the final images around $42 \mathrm{~au}$ (Fig.~\ref{fig:fig1_final_images}) has no strong imprints in the spectral index profiles. Then, even though higher than in the optically thick inner part, the values at $r > 50 \mathrm{~au}$ are small, especially in comparison to what was obtained by \citet[]{cieza2016imaging} (see solid grey line in the left panel of Fig.~\ref{fig:fig3_SpInd_profile}). In fact, they found values of about $\alpha \approx 3.5$ at radii $r > 50 \mathrm{~au}$. The difference can be explained by the fact that they only used data from Band 6 (see Sect.~\ref{sec:observations}), while we are computing our spectral indices with multiples bands, allowing us to probe a larger range of wavelengths and leading to a more robust estimate of the dust spectral index. This difference has important implications for the inferred particle sizes. \citet[]{schoonenberg2017pebbles} best reproduced this larger spectral index (grey line in the left panel of Fig.~\ref{fig:fig3_SpInd_profile}) with a model featuring the many-seeds scenario and the rapid re-coagulation of the grains into $300 \mathrm{~\mu m}$ aggregates. Instead, the lower value of the spectral index obtained here ($\alpha \approx 2.5$), as we will see further, implies the presence of considerably larger particles.

Individual profiles also display distinct features. The radial extent of the inner region characterised by $\alpha \lesssim 2$ varies with the wavelength, extending up to $30 \mathrm{~au}$ in the spectral index profile at $0.88-1.3 \mathrm{~mm}$, while reaching $10-20 \mathrm{~au}$ at $1.3-2.0 \mathrm{~mm}$ and $2.0-3.1 \mathrm{~mm}$. Due to the steep slope that follows the inner $\alpha \lesssim 2$ region, its exact radial extent may be affected by the absolute flux calibration systematic errors of ALMA. The value of the spectral index in the outer disc also differs, being slightly higher ($\alpha \approx 2.75$) at $1.3-2.0$ and $2.0-3.1 \mathrm{~mm}$. These wavelength-specific features can be used to characterise the properties of dust particles, such as the dust temperature or surface density, which will be the focus of future work. For $r > 125 \mathrm{~au}$, the spectral index at $1.3-2.0$ and $2.0-3.1 \mathrm{~mm}$ seems to decrease with the distance to the star. This could indicate the presence of larger aggregates, however, at such large distances, we argue that it might just be due to the larger errors.

Overall, our spectral index radial profiles suggest the presence of larger particles than previously assumed in V883 Ori \citep[$a \approx 300 \mathrm{~\mu m}$,][]{schoonenberg2017pebbles}. In fact, assuming an MRN size distribution, it would require maximum particle sizes in excess of $1 \mathrm{~cm}$ to reach spectral indices as low as $\alpha = 2.5$ in the optically thin regime \citep[e.g. Fig.4 in][]{testi2014dust}. However, fragmentation, radial transport, and the accretion outburst itself may significantly alter the size distribution from a standard MRN distribution \citep[]{brauer2008coagulation, birnstiel2011dust, houge2023collisional}. To further understand what our observational results imply concerning the physics of dust aggregates in outbursting discs, we will dedicate the next part of this manuscript to study two specific scenarios (many-seeds vs. resilient) with dust evolution simulations that solve self-consistently for the evolution of the entire dust size distribution. Note that obtaining a detailed fit of the observed spectral indices is beyond the scope of this paper, given the intricacies of all the different dust evolution models, and unknowns about disc parameters and initial conditions. Rather, we seek to model how the two end-member cases (many seeds vs. resilient) impact the dust distribution and evolution of the spectral index, to distinguish which scenario is the most appropriate to describe the dust content of V883 Ori. Specifically, our main goal is to identify which model can reproduce the newly-inferred low spectral index ($\alpha \approx 2.5$) in the optically thin region ($r>50 \mathrm{~au}$).

\begin{figure*}
    \includegraphics[width=\textwidth]{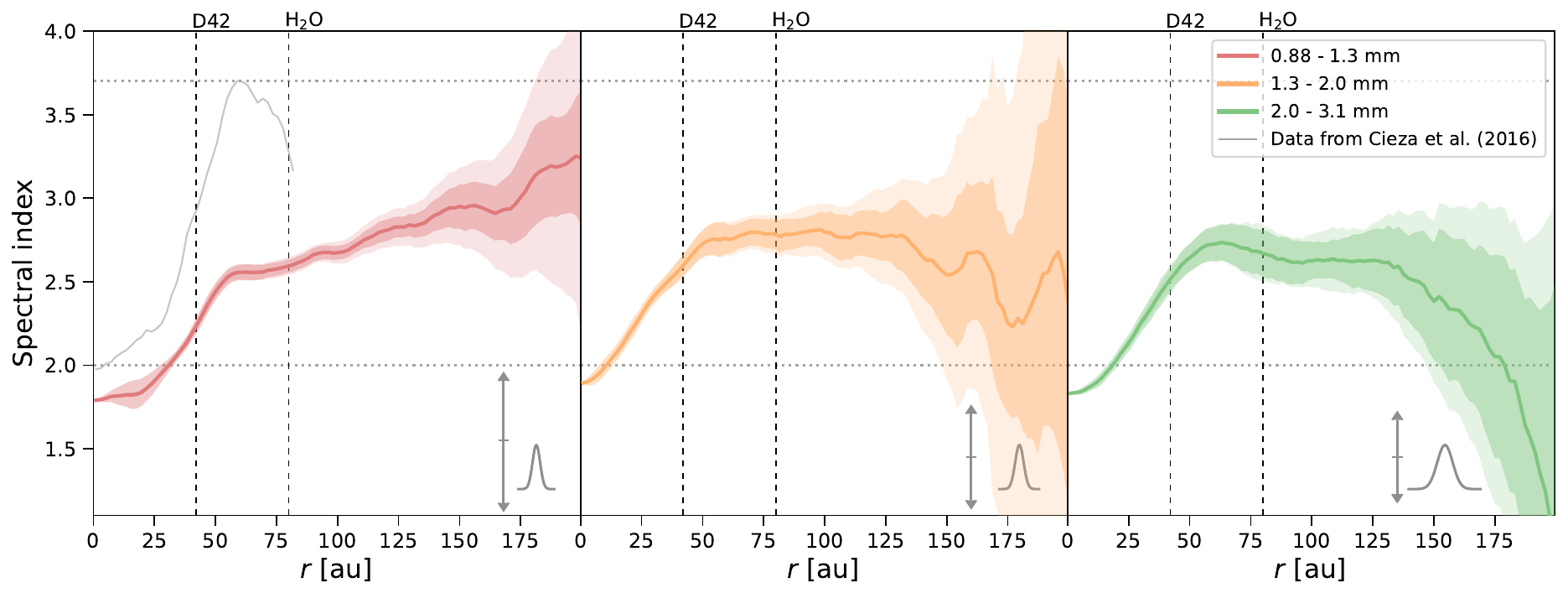}
    \caption{Spectral index radial profiles of the dust emission of V883 Ori at $0.88 - 1.3 \mathrm{~mm}$ (red curve), $1.3 - 2.0 \mathrm{~mm}$ (orange curve), and $2.0 - 3.1 \mathrm{~mm}$ (green curve). The profiles are retrieved by azimuthally averaging the spectral index maps obtained from the multi-bands images (Sect. \ref{sec:observations}). We also show the spectral index obtained from the Band 6 data {derived by} \citet[]{cieza2016imaging} (grey curve){. More information, including the error bars, can be found in the original paper}. The shaded areas are the error of the mean at each radii, and in fainter colours the azimuthal average of the spectral index error maps produced by \texttt{tclean}. The bottom right inset shows the geometric mean of the beam size for each spectral index map (Table~\ref{tab:final_images_info}) and the vertical grey lines represent the systematic error introduced by the flux calibration uncertainty. Horizontal dotted lines indicate $\alpha=2$ (optically thick emission in the Rayleigh–Jeans limit) and $\alpha=3.7$ \citep[typical value for micrometer dust grains in the ISM, see][]{finkbeiner1999extrapolation}. The vertical dashed lines represent respectively the dark annulus visible in Fig.~\ref{fig:fig1_final_images} and the water snowline position \citep[]{tobin2023deuterium}.}
    \label{fig:fig3_SpInd_profile}
\end{figure*}

\section{Modeling the dust evolution in V883 Ori}
\label{sec:simulation_methods}

To explore the evolution of the dust population in V883 Ori, we use the code \texttt{DustPy}\footnote{We used \texttt{DustPy} v1.0.3 for the simulations presented in this manuscript.} \citep[][]{stammler2022dustpy} which models the coagulation, fragmentation, and transport of dust particles in a 1D radial disc profile. In this Section, we describe our disc and dust models, and how we implemented an accretion outburst and the two sublimation scenarios into \texttt{DustPy}. We present the results of our dust evolution simulations in Sect.~\ref{sec:dustpy_results}.

\subsection{V883 Ori disc model}
\label{sec:dustpy_setup}

We modify the initial disc model of \texttt{DustPy} based on observational constraints of V883 Ori. V883 Ori is a T Tauri star of mass $1.3~\mathrm{M_\odot}$ \citep[][]{cieza2016imaging} surrounded by a disc of estimated mass $M_\mathrm{disc} \approx 0.3~\mathrm{M_\odot}$ \citep[][]{cieza2018alma, leemker2021chemically}. We assume the initial gas surface density profile to follow a tapered power-law \citep[]{lynden1974evolution, hartmann1998accretion}
\begin{equation}
\Sigma_\mathrm{g} (r) = \Sigma_\mathrm{c} \bigg( \dfrac{r}{r_c} \bigg)^{-\gamma} \mathrm{exp} \bigg[ -\bigg( \dfrac{r}{r_c} \bigg)^{2-\gamma}\bigg],
\label{eq:surface_density}
\end{equation}
where $r_\mathrm{c}=75 \mathrm{~au}$ \citep[][]{van2018methanol, leemker2021chemically}, $\gamma=1.25$, and 
\begin{equation}
\Sigma_\mathrm{c} = \dfrac{M_\mathrm{disc}(2-\gamma)}{2\pi r_c^2}.
\label{eq:surface_density_integrated}
\end{equation}

We simulate the disc over a discretized grid of $100$ radial cells {log-spaced} between $1\mathrm{~au}$ and $200~\mathrm{au}$. We initialize the simulation during the past quiescent phase of V883 Ori, where we assume the disc to be passively irradiated, characterised by a midplane temperature profile
\begin{equation}
\label{eq:quiescent_temperature_profile}
T_\mathrm{m}^\mathrm{qui} (r) = \left(\dfrac{\phi L_\mathrm{*}}{8 \pi\sigma_\mathrm{SB} r^2}\right)^{0.25},
\end{equation}
where $\phi=0.05$ is the disc opening angle and $L_\mathrm{*}=6~\mathrm{L_\odot}$ is the stellar luminosity \citep[]{cieza2016imaging}. Assuming the sublimation temperature of water to $150 \mathrm{~K}$ \citep[e.g.][]{collings2004laboratory, podolak2004note}, the quiescent water snowline is located at $\approx 3 \mathrm{~au}$.

\subsection{Dust model}
\label{sec:dustmodel}
We initialize the simulation with a population of grains following a size distribution characteristic of the ISM \citep[]{mathis1977size}, with a power-law $n(a) \propto a^{-3.5}$ from $0.5 \mathrm{~\mu m}$ monomers up to $100 \mathrm{~\mu m}$ aggregates \citep[assuming dust growth in the early stages of the disc evolution, e.g.][]{bate2022dust}. We adopt a dust composition following \citet[]{birnstiel2018disk}, where grains are made of a mixture of silicates, troilite, and refractory organics, with a relative mass abundance respectively set to $0.411$, $0.093$, and $0.496$. With a respective material density of $3.3$, $4.83$, and $1.50 \mathrm{~g~cm^{-3}}$, the overall bulk density of dust grains is $\rho_\mathrm{s} = 2.11 \mathrm{~g~cm^{-3}}$. Outside the water snowline, $\mathrm{H_2O}$ molecules are frozen out onto dust particles such that the water fraction is $20\%$ of the total mass, comparable to measurements of comet 67P/Churyumov-Gerasimenko \citep[]{patzold2016homogeneous}. Using a water ice density of $0.92 \mathrm{~g~cm^{-3}}$, it leads to icy dust grains with a bulk density $\rho_\mathrm{s} =  1.67 \mathrm{~g~cm^{-3}}$. We set the fragmentation velocity to $v_\mathrm{frag} = 1 \mathrm{~m~s^{-1}}$ inside the water snowline, and $10 \mathrm{~m~s^{-1}}$ outside, in agreement with laboratory experiments finding enhanced stickiness for water-rich dust particles \citep[e.g.][]{supulver1997sticking, gundlach2014stickiness}. Regarding the dust structure, we assume dust grains to coagulate following a compact aggregation model, remaining spherical and homogeneously mixed in dust and ice at all times.

\subsection{Outburst event}
\label{sec:otb_model}

We let the dust population evolve in quiescent conditions until $t_\mathrm{otb} = 10^5 \mathrm{~yr}$, marking the onset of the accretion outburst. This time, chosen to be long enough for aggregates to reach the fragmentation barrier but short enough for radial drift to not have removed a considerable amount of dust (see however Sect.~\ref{sec:dustpy_turbulence}), is comparable to the typical time span between outburst events \citep[]{contreras2019determining}, and, within a few factors, in agreement with the estimated age of V883 Ori \citep[]{cieza2016imaging}. After that time, we mimic an accretion outburst by raising the dust and gas temperature. We assume the outburst temperature profile to be constant throughout the accretion outburst, and rather than using Eq.~\ref{eq:quiescent_temperature_profile} with the current bolometric luminosity of V883 Ori, {we choose to fix the new outburst temperature profile to}
\begin{equation}
\label{eq:outburst_temperature_profile}
    T_\mathrm{m}^\mathrm{otb} (r) = 150\left( \dfrac{r}{80 \mathrm{~au}} \right)^{-0.5},
\end{equation}
{where the power-law index is identical to the one used in \citet{schoonenberg2017pebbles} and consistent with the one of a passively irradiated disc \citep[]{kenyon1987spectral}; and the normalisation has been updated so that the water snowline position corresponds roughly to $80 \mathrm{~au}$, as suggested by\footnote{We assume the sublimation temperature of water to remain at $T=150 \mathrm{~K}$, while in reality it would slightly decrease in the outer disc due to the lower gas density \citep[e.g.][]{supulver2000formation}.} \citet{tobin2023deuterium}.} We discuss the assumption of a constant temperature profile in the context of outburst decay and variability in Sect.~\ref{sec:discussion_variability}.} We assume the temperature to increase instantaneously to the outburst value, as the heating timescale of the dust is generally smaller than the typical outburst duration \citep[][]{johnstone2013continuum}. We further assume the disc vertical structure to instantaneously adapt to a new hydrostatic equilibrium when the temperature is modified. This is a valid assumption when the gas thermal timescale \citep[e.g. Equation 7 in][]{ueda2021thermal} is short compared to the dynamical timescale and the outburst duration, which is the case in our disc model at all relevant radii.

The onset of the accretion outburst in V883 Ori was not directly observed, but based on observations of its reflection nebula in 1888 by \citet[][]{pickering1890detection}, \citet[][]{strom1993discovery} argued that it has been active for at least $135 \mathrm{~yr}$, and likely began even before that observation. Due to the uncertainty regarding the exact onset time, we keep the outburst active for $10^4 \mathrm{~yr}$. {This duration is considerably longer than what is expected for FUor-type accretion outbursts \citep[]{audard2014episodic}, but is chosen to allow us to study the evolution of dust particles on longer timescales as they readjust to the new coagulation/fragmentation equilibrium (see Sect.~\ref{sec:dustpy_results}).} We end the simulation after that time ($t_\mathrm{otb}+10^4 \mathrm{~yr}$), and do not model the post-outburst phase and the recondensation of water vapor \citep[for that matter, see][]{vorobyov2022evolution, houge2023collisional}.

\subsection{Dust response to sublimation}
\label{sec:methods_dustreponse_sublimation}

In the midplane between the initial quiescent snowline ($3 \mathrm{~au}$) and outburst water snowline ($80 \mathrm{~au}$), water is initially in ice phase deposited onto the surface of refractory particles, before sublimating when the outburst starts at $t = t_\mathrm{otb}$. This zone will be further referred to as the sublimation region. Similarly to previous works \citep[]{cieza2016imaging, schoonenberg2017pebbles}, we assume water ice to sublimate instantaneously at the outburst onset. It is a valid assumption in warm outbursting conditions, as the desorption of volatile molecules is very sensitive to the temperature {\citep[e.g.][]{supulver2000formation}}. Moreover, in V883 Ori, water vapor and signs of an active water chemistry with HCO+ is observed up to $100 \mathrm{~au}$ \citep[]{leemker2021chemically, tobin2023deuterium}, providing additional evidence that dust particles up to such radii underwent heating and sublimation. 

In the sublimation region, the composition of dust particles is modified at the outburst onset, leading to a change {in bulk density $\rho_\mathrm{s}$ and} in the fragmentation velocity $v_\mathrm{frag}$, decreasing from $10$ to $1 \mathrm{~m~s^{-1}}$ (Sect.~\ref{sec:dustmodel}). In addition, even in the absence of collisions, the sublimation of water ice may lead directly to the disintegration of icy dust aggregates (see Sect.~\ref{sec:introduction}). Laboratory experiments by \citet[]{aumatell2011breaking} investigated the sublimation of mm-sized pure ice pebbles, and found that they break up in a large number ($\approx 10^4$) of smaller pieces, supporting the idea that sublimation disrupts icy dust aggregates into many $\mu \mathrm{m}$-sized silicate grains. More recently, however, \citet[]{spadaccia2022fate} studied mm-sized pebbles made of a mix of dust and ice and found that they are able to survive sublimation, especially if the ice content is around $15\%$ of the pebble mass, which is around what we expect for icy particles in protoplanetary discs (see Sect.~\ref{sec:dustmodel}). In agreement with their findings, we implement two scenarios for the response of icy dust aggregates to sublimation at the outburst onset: (a) the many-seeds scenario, where we disrupt all dust particles in the sublimation region back to the monomer size ($0.5 \mathrm{~\mu m}$); and the resilient scenario, where particles survive and are left unchanged. {Note that in the resilient scenario, individual particles masses would also decrease by a fraction relative to their ice content ($20\%$ in our dust model, see Sect.~\ref{sec:dustmodel}), but as shown in \citet[]{houge2023collisional}, it is the compositional change and subsequent lowered fragmentation velocity that dominates the appearance and behaviour of resilient aggregates. Furthermore}, we do not extend the many-seeds model to the general case of pebbles drifting through the water snowline at any time of the simulation. At the snowline location, we therefore keep the standard implementation of \texttt{DustPy}, where pebbles survive as they drift through the snowline, but quickly fragment as they become less resistant to fragmentation.

\subsection{Dust evolution results}
\label{sec:dustpy_results}

\subsubsection{Many-seeds scenario}
\label{sec:results_manyseeds}

\begin{figure*}
    \includegraphics[width=\textwidth]{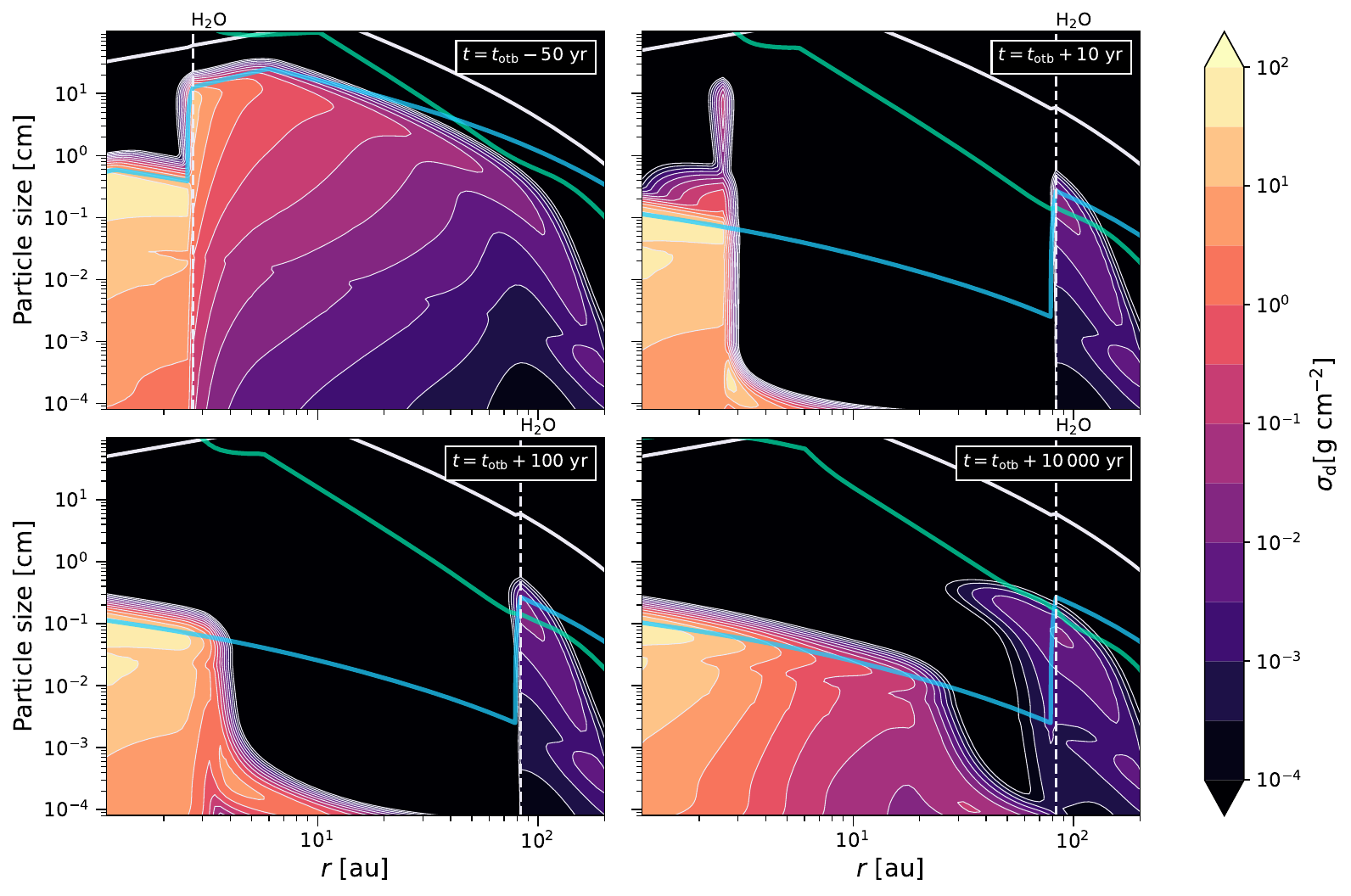}
    \caption{Dust density distribution as a function of particle size in the many-seeds scenario. The colormap refers to the surface density of dust in a given size range at a given radius. The four panels indicate different key times of the simulation: before the outburst (top-left), after an outburst duration of $10 \mathrm{~yr}$ (top-right), after an outburst duration of $100 \mathrm{~yr}$ (bottom-left), and after an outburst duration of  $10\,000 \mathrm{~yr}$ (bottom-right). The vertical white dotted line ($\mathrm{H_2O}$) specifies the current snowline position. The blue and green solid lines respectively represent the fragmentation and radial drift growth barriers. The solid white line indicates the size at which particles have a Stokes number $\mathrm{St}=1$.}
    \label{fig:Fig4_MS_DustPy}
\end{figure*}

We begin with the results from the many-seeds scenario. We present in Fig.~\ref{fig:Fig4_MS_DustPy} the dust density distribution as a function of particle size at different key times of the simulation: right before the outburst (top-left); after an outburst duration of $10 \mathrm{~yr}$ (top-right); after an outburst duration of $100 \mathrm{~yr}$, which corresponds approximately to the minimum duration of the outburst in V883 Ori following the measurements of \citet[]{pickering1890detection} (bottom-left); and after an outburst duration of $10^4 \mathrm{~yr}$ (bottom-right). {As mentioned in Sect.~\ref{sec:otb_model}, the latter duration is much longer than what is typically expected for an FUor-type outburst ($\approx 100 \mathrm{~yr}$), but is included to allow us to discuss how dust readjust to the new coagulation/fragmentation equilibrium.} The blue and green lines in each panel in Fig.~\ref{fig:Fig4_MS_DustPy} represent respectively an estimate of the current fragmentation and radial drift barriers, after which dust growth through coagulation is halted \citep[]{birnstiel2012simple}.

Before the outburst (top-left panel in Fig.~\ref{fig:Fig4_MS_DustPy}), the dust population has evolved for $10^5 \mathrm{~yr}$ and reached the fragmentation or radial drift barrier up to $\approx 100 \mathrm{~au}$, reaching sizes $\geq 1 \mathrm{~cm}$. When the outburst starts (top-right panel in Fig.~\ref{fig:Fig4_MS_DustPy}), water 
sublimates in the sublimation region ($3 < r < 80 \mathrm{~au}$), upon which all dust particles fall apart to $\mathrm{\mu m}$-sized silicate grains, in agreement with the many-seeds scenario (Sect.~\ref{sec:methods_dustreponse_sublimation}). Due to the change in composition in the sublimation region, the fragmentation barrier (solid blue line) decreases by about two orders of magnitude, as $a_\mathrm{max} \propto v_\mathrm{frag}^2$ \citep[]{birnstiel2011dust}. The fragmentation barrier also slightly decreases throughout the disc due to the rise in temperature, as it results in larger relative velocities between dust particles \citep[$a_\mathrm{max} \propto T^{-1}$, see][]{birnstiel2011dust}. This temperature-induced fragmentation can be seen in the inner disc ($r < 3 \mathrm{~au}$) on the top-right panel of Fig.~\ref{fig:Fig4_MS_DustPy}, where pebbles that are now too large (i.e. above the blue line) fragment to reach the new fragmentation barrier. Thanks to high density in the inner disc and the small amplitude of the size change required, it only takes a few years for the inner disc population to adapt to the new coagulation/fragmentation equilibrium. Note that these particles did not disrupt to $\mathrm{\mu m}$-sized grains because they are located inside the quiescent water snowline, where water is in a vapor state at all times. We can also see a thin patch of large particles (${>}1 \mathrm{~cm}$) directly at the quiescent snowline, corresponding to pebbles that just drifted through the snowline before the outburst started, and have not yet fragment. The radial drift barrier (solid green line) also decreases by factor of a few, which only impacts particles outside of the outburst snowline ($r > 80 \mathrm{~au}$). 

After $100 \mathrm{~yr}$ past the outburst onset (bottom-left panel in Fig.~\ref{fig:Fig4_MS_DustPy}), we notice that dust aggregates are re-coagulating in the sublimation region. However, the collision rates decrease with increasing radial distance, which leads to a growth front propagating outward with time since the outburst onset \citep[see also][]{vorobyov2022evolution}. After $100 \mathrm{~yr}$, it is located at $\approx 5\mathrm{~au}$: dust particles inside of that radius already reached the new fragmentation barrier, while particles outside of that radius are still smaller than $10 \mathrm{~\mu m}$. Even after $10^4 \mathrm{~yr}$ (bottom-right panel in Fig.~\ref{fig:Fig4_MS_DustPy}), the dust population has not reached the new fragmentation barrier outside of $20\mathrm{~au}$. After such extended outburst duration, we also start to notice the impact of radial drift, replenishing the region between $30-80 \mathrm{~au}$ in mm-sized particles although most of the dust mass in this region is still stored in grains with sizes below $10 \mathrm{~\mu m}$. 

Overall, in the many-seeds scenario, $\mathrm{\mu m}$-sized grains require considerable timescales to re-coagulate, and no dust aggregates over a millimetre size can be found in the sublimation region ($3 < r < 80 \mathrm{~au}$) shortly after the outburst. The only exception comes from pebbles drifting through the outburst water snowline ($r > 80 \mathrm{~au}$), but these only appear after ${\approx} 10^4 \mathrm{~yr}$.

\subsubsection{Resilient scenario}
\label{sec:results_resilient}

\begin{figure*}
    \includegraphics[width=\textwidth]{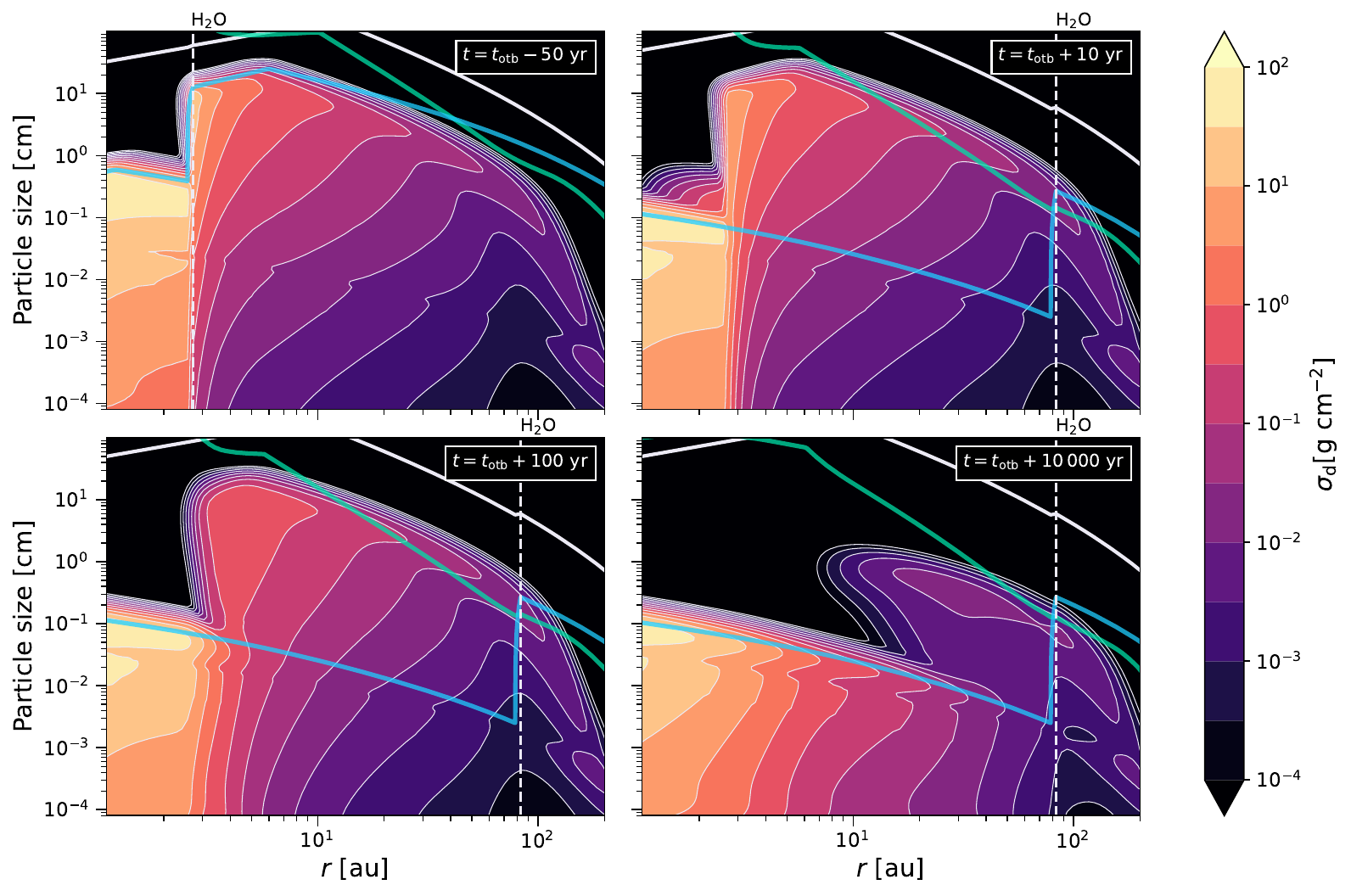}
    \caption{Dust density distribution as a function of particle size in the resilient scenario. The colormap refers to the surface density of dust in a given size range at a given radius. The four panels indicate different key times of the simulation: before the outburst (top-left), after an outburst duration of $10 \mathrm{~yr}$ (top-right), after an outburst duration of $100 \mathrm{~yr}$ (bottom-left), and after an outburst duration of  $10\,000 \mathrm{~yr}$ (bottom-right). The vertical white dotted line ($\mathrm{H_2O}$) specifies the current snowline position. The blue and green solid lines respectively represents the fragmentation and radial drift growth barriers. The solid white line indicates the size at which particles have a Stokes number $\mathrm{St}=1$.}
    \label{fig:Fig5_Res_DustPy}
\end{figure*}

We now present in Fig.~\ref{fig:Fig5_Res_DustPy} the dust density distribution as a function of particle size in the resilient scenario, at identical times as Fig.~\ref{fig:Fig4_MS_DustPy}. The pre-outburst distribution is identical to Fig.~\ref{fig:Fig4_MS_DustPy}, with the dust growth halted by the fragmentation or radial drift barrier up to $\approx 100 \mathrm{~au}$. When the outburst starts (top-right panel in Fig.~\ref{fig:Fig5_Res_DustPy}), {the resilient scenario (Sect.~\ref{sec:methods_dustreponse_sublimation}) assumes dust aggregates survive sublimation and the dust distribution is therefore similar to the pre-outburst state. However, as the composition changed to dry silicates, the fragmentation barrier has decreased by two orders of magnitude} \citep[$a_\mathrm{max} \propto v_\mathrm{frag}^2$, see][]{birnstiel2011dust}. Pebbles that are now too large, i.e. above the blue line, start fragmenting, trying to settle in the new (local) coagulation/fragmentation equilibrium. Nevertheless, it requires time to fragment all the pebbles, and after $100 \mathrm{~yr}$ (bottom-left panel in Fig.~\ref{fig:Fig5_Res_DustPy}), fragmentation still dominates. Similarly to the many-seeds scenario, the dust population responds to the outburst inside-out, adapting faster closer to the host star thanks to larger dust densities and collision rates. It leads to a fragmentation front propagating outward with time since the outburst onset\footnote{Note that efficient fragmentation can even extend in the post-outburst phase, {with the largest pebbles remaining} relatively dry as water re-condenses preferentially on the smaller grains \citep[see][]{houge2023collisional}.}. Even after $10^4 \mathrm{~yr}$ (bottom-right panel in Fig.~\ref{fig:Fig5_Res_DustPy}), pebbles are still fragmenting and a considerable amount of dust with sizes above the local fragmentation barrier (i.e. above the blue line) persists at radii between $\approx 10$ and $80 \mathrm{~au}$. 

In summary, at several tens of au, where the collision timescales are fairly long as compared to the outburst typical duration ($\approx 100 \mathrm{~yr}$), the evolution of the dust size distribution during an outburst is set primarily by how individual icy dust aggregates respond to ice sublimation at the outburst onset.

\section{Comparisons between observations and dust simulations}
\label{sec:dustpy_spectralindex}

In this section, we convert the results from our dust evolution models into spectral indices, and compare them with the observed values (Fig.~\ref{fig:fig3_SpInd_profile}). Given the dust composition detailed in Sect.~\ref{sec:dustmodel}, we take the optical constants from \citet[]{draine2003interstellar} for astronomical silicates, \citet[]{henning1996dust} for troilite and refractory organics, and \citet[]{warren2008optical} for water ice. We then compute the mixed dielectric function using the Bruggeman effective medium theory {\citep[]{bruggeman1935berechnung}}. Next, we compute the absorption opacity and scattering opacity\footnote{We include the correction for the anisotropic scattering by multiplying the scattering opacity by the factor $(1-g_\nu)$, where $g_\nu$ is the forward scattering parameter.} using the \texttt{DSHARP-OPAC} package from \citet[]{birnstiel2018disk}. Finally, using the dust density distribution at each radius (self-consistently derived from our dust evolution models), we calculate the total absorption opacity $\kappa_\nu^\mathrm{abs}$ and total scattering opacity $\kappa_\nu^\mathrm{sca}$. From the total absorption and scattering opacity, we can calculate the dust emission intensity $I_\nu$ at each radius assuming a 1D slab model following \citep[]{sierra2019analytical, carrasco2019radial}
\begin{equation}
\label{eq:intensity}
I_\nu=B_\nu(T_{\mathrm{d}})[(1-\exp(-\tau_\nu / \mu))+\omega_\nu F(\tau_\nu, \omega_\nu)],
\end{equation}
where $T_\mathrm{d}$ is the {midplane} dust temperature, $B_\nu (T_{\mathrm{d}})$ is the black body emission at a frequency $\nu$ for a temperature $T_\mathrm{d}$, $\tau_\nu=\Sigma_\mathrm{d}(\kappa_\nu^\mathrm{abs} + \kappa_\nu^\mathrm{sca})$ is the optical depth, $\Sigma_\mathrm{d}$ is the dust surface density, $\omega_\nu = \kappa_\nu^\mathrm{sca} / (\kappa_\nu^\mathrm{sca} + \kappa_\nu^\mathrm{abs})$ is the dust albedo, $\mu = \cos{i}$, and $i=38.3^\circ$ is the inclination angle of V883 Ori (Sect.~\ref{sec:obs_results_spectral_index}). The flux $F(\tau_\nu, \omega_\nu)$ is itself obtained as \citep[]{sierra2019analytical, carrasco2019radial}
\begin{equation}
\label{eq:flux}
\begin{aligned}
F\left(\tau_\nu, \omega_\nu\right)= & \frac{1}{\exp \left(-\sqrt{3} \epsilon_\nu \tau_\nu\right)\left(\epsilon_\nu-1\right)-\left(\epsilon_\nu+1\right)} \\
& \times\left[\frac{1-\exp \left(-\left(\sqrt{3} \epsilon_\nu+1 / \mu\right) \tau_\nu\right)}{\sqrt{3} \epsilon_\nu \mu+1}\right. \\
& \left.+\frac{\exp \left(-\tau_\nu / \mu\right)-\exp \left(-\sqrt{3} \epsilon_\nu \tau_\nu\right)}{\sqrt{3} \epsilon_\nu \mu-1}\right],
\end{aligned}
\end{equation}

where $\epsilon_\nu = \sqrt{1-\omega_\nu}$. {This 1D slab model is a good approximation for protoplanetary discs, given that, at sub-mm wavelengths, we mainly trace large pebbles highly settled into the disc midplane. From Eq.~\ref{eq:intensity}, we derive the dust emission spectral index as $\alpha_\mathrm{\nu_1 - \nu_2} = \ln{(I_{\nu_1}/I_{\nu_2})}/\ln{(\nu_1/\nu_2)}$, where $\nu_1$ and $\nu_2$ are chosen in agreement with the central wavelength of the images at each band (Table \ref{tab:final_images_info}). Note that the spectral index is computed from our dust evolution models, sampled on the log-spaced radial grid described in Sect.~\ref{sec:dustpy_setup}. As a result, the spectral index from our models is not convolved with the resolution of the ALMA images, which would only affect our results at relatively small spatial scales given the high angular resolution of the observations (see Table~\ref{tab:final_images_info}).
}

\begin{figure*}
    \includegraphics[width=\textwidth]{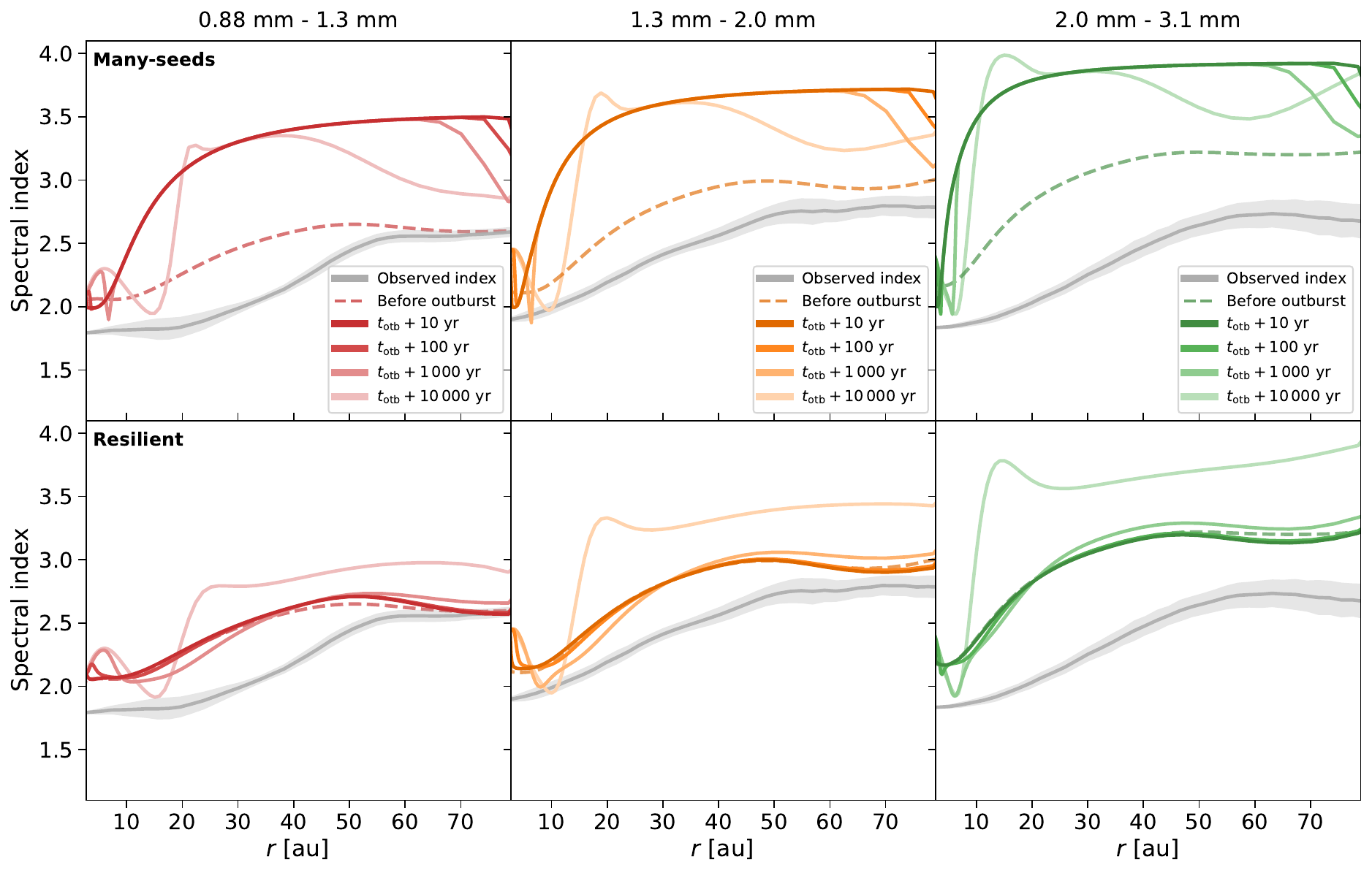}
    \caption{Spectral index radial profiles in the many-seeds scenario (upper panels) and resilient scenario (lower panels) after different outburst duration at $0.88 - 1.3 \mathrm{~mm}$ (red curves), $1.3 - 2.0 \mathrm{~mm}$ (orange curves), and $2.0 - 3.1 \mathrm{~mm}$ (green curves). We include at each wavelength the pre-outburst spectral index as a dashed line, and the observed spectral index with a solid grey line (Fig.~\ref{fig:fig3_SpInd_profile}).}
    \label{fig:Fig4_DustPy_SpIndex}
\end{figure*}

We begin with the spectral index radial profiles of the dust evolution model in the many-seeds scenario, presented in the upper panels of Fig.~\ref{fig:Fig4_DustPy_SpIndex} at different times before and after the outburst onset. Following the goal of this work, we focus on the sublimation region of V883 Ori ($3 < r < 80 \mathrm{~au}$). The pre-outburst spectral index is relatively close to the observed data, featuring an optically thick inner disc ($r < 15 \mathrm{~au}$) with values around $\alpha \approx 2$, followed at larger radii by optically thin emissions with a spectral index $\alpha \approx 2.5$ at $0.88-1.3 \mathrm{~mm}$. The relatively low spectral index is due to the presence of large cm-sized pebbles in these regions during the pre-outburst phase (Fig.~\ref{fig:Fig4_MS_DustPy}). Comparing different wavelengths, the shape of the pre-outburst profile is similar, although it shifts to slightly higher spectral indices at larger wavelengths due to varying optical depth and albedo. The model does not manage to reproduce the values $\alpha<2$ in the inner disc, which could be caused by unknowns about the disc parameters, or by the fact that we do not consider a vertical temperature profile, which was shown to decrease the spectral index in the optically thick regime \citep[]{sierra2020effects}. When the outburst starts, dust particles disrupt to $\mathrm{\mu m}$-sized grains, and the spectral indices increase significantly, reaching values close to $\alpha = 3.7$, similar to small grains in the ISM \citep[]{finkbeiner1999extrapolation}. With time, the spectral index begins to evolve in the innermost and outermost part of the sublimation region. This is due respectively to the re-coagulation of dust particles (see growth front in Sect.~\ref{sec:results_manyseeds}) and the inward drift of pebbles that survived outside of the outburst water snowline ($r>80\mathrm{~au}$). However, these effects are not sufficient to replenish in large particles the sublimation region of V883 Ori within the typical range of outburst duration \citep[see also][for a discussion on drift timescale and outburst duration]{schoonenberg2017pebbles}.

We then present the spectral index radial profiles of the resilient scenario in the lower panels of Fig.~\ref{fig:Fig4_DustPy_SpIndex}. The pre-outburst spectral index is identical to the many-seeds case. However, when the outburst starts, the spectral index remains relatively close to the pre-outburst profile, much closer to the observed data (Fig.~\ref{fig:fig3_SpInd_profile}), with values $\alpha \approx 2.5$. Over time, more pebbles are fragmenting causing the spectral index to vary and move away from the observed data. This is particularly true in the inner part of the sublimation region (see fragmentation front in Sect.~\ref{sec:results_resilient}). In our dust evolution models, the resilient scenario is therefore the most appropriate to describe the dust population of V883 Ori.

\subsection{Varying the turbulence parameter}
\label{sec:dustpy_turbulence}

\begin{figure*}
    \includegraphics[width=0.8\textwidth]{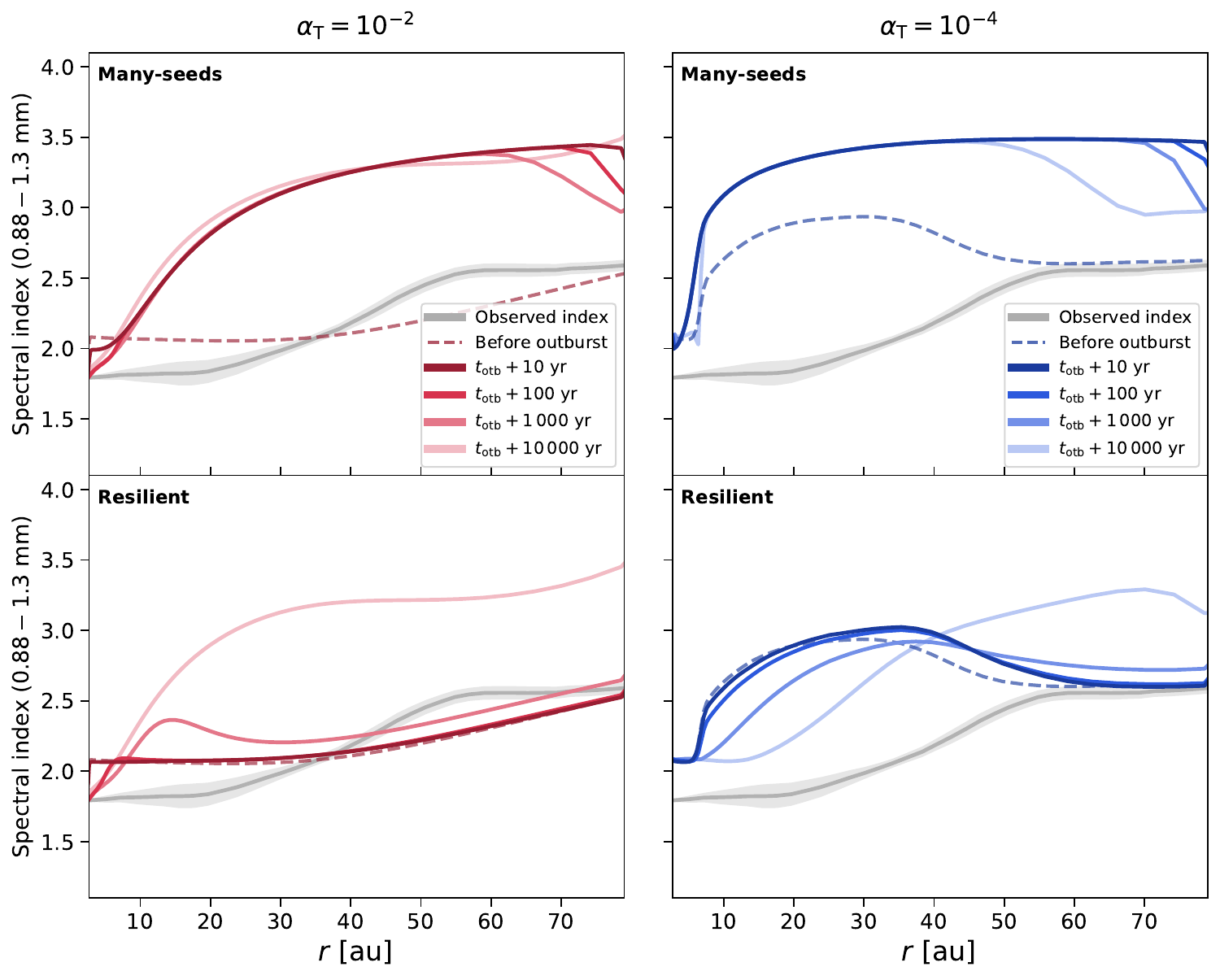}
    \caption{Spectral index radial profiles in the many-seeds scenario (upper panels) and resilient scenario (lower panels) after increasing outburst duration at $0.88 - 1.3 \mathrm{~mm}$ for $\alpha_\mathrm{T}=10^{-2}$ (left panels, red curves) and $\alpha_\mathrm{T}=10^{-4}$ (right panels, blue curves). We include the pre-outburst spectral indices as dashed lines, and the observed spectral index with a solid grey line (Fig.~\ref{fig:fig3_SpInd_profile}).}
    \label{fig:Fig6_SpIndex_Turbulence}
\end{figure*}

As discussed in Sect.~\ref{sec:obs_results_spectral_index}, the spectral index and its temporal evolution are highly sensitive on uncertain disc and dust parameters. Notably, the turbulence parameter $\alpha_\mathrm{T}$ is rather difficult to estimate in protoplanetary discs \citep[see][and references therein]{rosotti2023empirical}, but has a great impact on dust evolution: stronger turbulence leads to higher relative velocities between particles, which shifts the fragmentation barrier to smaller sizes $a_\mathrm{max}$, reducing the speed of radial drift \citep[]{birnstiel2011dust}. Moreover, high relative velocities result in higher collision rates, so that coagulation proceeds more rapidly. To investigate the impact of the choice of $\alpha_\mathrm{T}$ on our results, we perform additional dust evolution simulations using $\alpha_\mathrm{T}=10^{-2}$ and $\alpha_\mathrm{T}=10^{-4}$. We present their spectral index radial profile at $0.88 - 1.3 \mathrm{~mm}$ in Fig.~\ref{fig:Fig6_SpIndex_Turbulence}, and refer the interested reader to Appendix \ref{sec:appendix_1} for more details about how the turbulence influences the dust density distribution. 

When comparing the pre-outburst spectral index for different turbulence values (dashed red and dashed blue curves in Fig.~\ref{fig:Fig6_SpIndex_Turbulence}), we notice the significant impact of the turbulence parameter on the spectral index radial profile. For example, for lower turbulence $\alpha_\mathrm{T}=10^{-4}$ (right panels in Fig.~\ref{fig:Fig6_SpIndex_Turbulence}), pebbles are larger and more efficiently transported, preventing them to fragment and redistribute mass in smaller particles. When the outburst starts at $t_\mathrm{otb} = 10^5 \mathrm{~yr}$, the reservoir of mm-sized particles has significantly decreased, leading to a larger spectral index inside $50 \mathrm{~au}$ (see Appendix \ref{sec:appendix_1} for more details). However, when comparing the resilient and many-seeds scenarios for identical turbulence values (top vs. bottom panels in Fig.~\ref{fig:Fig6_SpIndex_Turbulence}), we reach similar conclusions as in the previous section: the spectral index in the many-seeds case is too high, and does not well represent the observed dust population in V883 Ori (Fig.~\ref{fig:fig3_SpInd_profile}).

To conclude, based on comparisons between observations (Fig.~\ref{fig:fig3_SpInd_profile}) and dust evolution models (Fig.~\ref{fig:Fig4_DustPy_SpIndex} and Fig.~\ref{fig:Fig6_SpIndex_Turbulence}), we find that large dust particles must be present in V883 Ori to explain the observed spectral indices. Given that, in the many-seeds scenario, mm-sized aggregates require timescale significantly larger than the typical outburst duration to re-appear, the only way to explain the dust population in V883 Ori is that the resilient scenario applies, hence that dust particles survived the sublimation of their water ice content at the onset of the accretion outburst.

\section{Discussion}
\label{sec:discussion}

\subsection{How does dust respond to an accretion outburst?}
\label{sec:discussion_dust_sublimation}

{
Previous works have studied the dust size distribution in detail in two protoplanetary discs undergoing accretion outbursts: FU Ori and V883 Ori. For FU Ori, \citet[]{liu2021millimeter} employed multi-wavelength observations coupled with MCMC analysis to infer the presence of large pebbles (${>}1\mathrm{~mm}$) surviving the water-sublimating temperature, in line with the resilient scenario. For V883 Ori, however, based exclusively on ALMA Band 6 data obtained by \citet[]{cieza2016imaging}, \citet[]{schoonenberg2017pebbles} best reproduced the observed spectral index by assuming the fast re-coagulation of refractory dust grains after their total disruption to $\mu$m size at the outburst onset, i.e. the many-seeds scenario.} 

In this paper, thanks to new multi-wavelength analysis of V883 Ori in Band 7, 6, 4, and 3 (see Sect.~\ref{sec:observational_results}), we argue instead that the dust particles in V883 Ori also survived the sublimation event at the outburst onset. With the same scenario fitting FU Ori and V883 Ori, we conclude that icy dust aggregates follow the resilient scenario, and survive to water sublimation during accretion outbursts. As discussed in \citet[]{houge2023collisional}, these results imply that the impact of accretion outbursts on the dust size distribution is limited, and will mostly concern the composition of pebbles. In fact, the largest pebbles will require considerable time after the outburst to retrieve their pre-outburst water content {(between $500$ and $2000 \mathrm{~yr}$ at $5\mathrm{~au}$, see Fig.~13 in \citealt{houge2023collisional}). These timescales are generally longer than the time required for a cloud of pebbles to gravitationally collapse into planetesimals \citep[${\sim}100 \mathrm{~yr}$, see][]{visser2021radial}. Therefore, if the formation of planetesimals is triggered by the outburst, it may lead to the formation of water-poor planetesimals. Note, however, that the collapse of a pebble cloud first requires the formation of a highly concentrated clump of material \citep[]{johansen2007rapid, johansen2009particle}, e.g. induced by the streaming instability \citep[]{youdin2005streaming}, which itself takes hundreds of orbital periods to form \citep[]{carrera2015form}. It should thus be already present for the outburst to have an impact on planetesimal formation.}

We note that in this work, we adopt two end-member scenarios: (1) the many-seeds case, where all aggregates fall apart, and (2) the resilient case, where all aggregates survive. However, depending on the complexity of the pre-outburst size distribution, a hybrid scenario can be envisioned. For example, \citet[]{spadaccia2022fate} found that the ability of icy dust aggregates to survive sublimation depends on their water mass fraction, and on the mineralogy and size distribution of the grains they are made of. One could imagine these parameters to vary within the dust population, leading to an hybrid scenario which could be studied if based on additional inputs from laboratory experiments. However, to explain the relatively low values of the spectral indices we measure in V883 Ori (Fig.~\ref{fig:fig3_SpInd_profile}), cm-sized pebbles must still survive sublimation. Whether or not smaller grains follow the same trend would not impact our findings and conclusions.

\subsection{What about pebbles drifting through the water snowline?}
\label{sec:discussion_drift_snowline}

We can compare the behaviour of icy aggregates in the sublimation region of outbursting discs to that of pebbles drifting across the (static) water snowline in a quiescent disc. Accretion outbursts represent an important stress for dust particles, as the temperature changes abruptly to higher values ($\gtrsim 100 \mathrm{~K}$ in a few years). In comparison, the temperature of pebbles drifting through the water snowline in a quiescent disc increases slowly with the inward radial motion. We can express the heating rate of drifting pebbles as
\begin{equation}
\label{eq:heating_rate}
    \dfrac{\mathrm{d}T}{\mathrm{d}t}= \dfrac{\mathrm{d}T}{\mathrm{d}r} v_\mathrm{r},
\end{equation}
where $v_\mathrm{r}$ is the radial drift velocity. Considering the upper limit case of pebbles characterised by $\mathrm{St}=1$, for which the radial drift velocity reaches the maximum value, and using Eq.~\ref{eq:quiescent_temperature_profile} for the temperature profile in a quiescent disc, we find $\mathrm{d}T/\mathrm{d}t < 1 \mathrm{~K~yr^{-1}}$ at the snowline location ($3 \mathrm{~au}$), representing a considerably less abrupt heating than an accretion outburst. For the drift scenario, \citet[]{piso2015c} found that, as a result of the shallow temperature profile, the desorption timescale of $\mathrm{H_2O}$ ice mantles near the water snowline can even be longer than the drift timescale, suggesting aggregates can traverse a significant distance before shedding their mantles. Based on the gentler heating and sublimation process, we speculate that if pebbles survive sublimation in an outbursting disc, they should also survive when drifting through the water snowline. 

The survival of pebbles drifting through the water snowline may have important implications for the formation of dry planetesimals interior to the snowline. In fact, the appearance of the traffic jam effect at that location either requires (Sect.~\ref{sec:introduction}): (1) that ice-covered particles are more resistant to fragmentation than bare silicates, which has been contested by recent laboratory experiments \citep[e.g.][]{gundlach2018tensile, musiolik2019contacts}; or (2) that dust particles fall apart to $\mu$m-sized monomers, which seems not to be the case as suggested by our findings. As a consequence, if future investigations confirm that ice-rich particles do not benefit from enhanced resistance to fragmentation (e.g. using the approach discussed in Sect.~\ref{sec:discussion_dustsubstructures}), then the formation pathway of dry planetesimals interior to the water snowline may need to be revised. In addition, without the release of many $\mu$m-sized grains able to diffuse through the snowline and boost the solids-to-gas ratio, the formation of icy planetesimals exterior to the snowline may also be lowered \citep[]{saito2011planetesimal, ida2016radial, schoonenberg2017planetesimal, hyodo2019formation, hyodo2021planetesimal}.

\subsection{Implications for the internal structure of dust aggregates}
\label{sec:discussion_dust_structure}

The fact that dust aggregates survive the sublimation of their water content can give us constrains about the actual structure of dust particles in protoplanetary discs, and how water ice is distributed on their surface. For icy porous aggregates, it is generally assumed that water ice covers the entire surface of monomers, bonding them to each other \citep[e.g.][]{okuzumi2012rapid, krijt2016tracing}. However, this approach contradicts our findings. In fact, if water ice glues refractory monomers to each other, then the sublimation of water ice should lead to a disruption of the aggregates, i.e. our many-seeds scenario. Finding large surviving pebbles in V883 Ori and FU Ori teaches us that: either (1) water ice does not cover the entire surface of monomers, which has important consequences for the collisional properties of porous aggregates, by changing the values of key parameters like the fragmentation limit $v_\mathrm{frag}$ or the rolling energy $E_\mathrm{roll}$, or (2) water ice does cover the surface of monomers, but, after sublimation, monomers are able to immediately collide and reconnect the aggregate. The latter could be tested with N-body simulations or laboratory experiments.

Alternatively, our findings concerning the resilient scenario could indicate that icy dust particles are rather made of a large silicate skeleton embedded in a water ice matrix, which is released at the outburst onset, as sublimation should not change the structure of the dust within the matrix. In that case, we speculate that the large dust skeleton would need to form before being covered in ice, meaning that dust particles primarily grow either: 1) in the disc atmosphere, 2) during the warm Class 0 phase {\citep[]{steinacker2010direct, ormel2011dust}}, or 3) inside the water snowline, before being redistributed to the outer disc. For the latter, we note that there are evidences that efficient outward transport should have occurred in the solar nebula, in order to explain the distribution of CAIs in Solar System objects \citep[e.g.][]{ciesla2007outward, marschall2023inflationary}. 

\subsection{Dust emission substructures in outbursting discs}
\label{sec:discussion_dustsubstructures}
Our findings have implications for the characterisation of outbursting discs using dust emission alone. First, the resilient scenario complicates the viability of using dust emission to detect the outburst water snowline. In fact, in the many-seeds scenario, we could have observed a sharp transition at the location of the snowline, as, inside, all particles fell apart to $\mu$m grains, while outside, water ice remained stable and large pebbles survived (Fig.~\ref{fig:Fig4_MS_DustPy}). In the favoured resilient case, however, we do not expect such a transition (Fig.~\ref{fig:Fig5_Res_DustPy}). In addition, models predicting the signature of the water snowline based on the pile-up of pebbles at that location, such as \citet[]{banzatti2015direct}, cannot readily be applied to the snowline of outbursting systems, as the timescale for dust pile-up is orders of magnitude higher than the outburst duration \citep[]{schoonenberg2017pebbles}.

Second, in the resilient scenario, we discussed in Sect.~\ref{sec:results_resilient} that the fragmentation of pebbles occurs inside-out from the quiescent snowline, propagating outward with time. As fragmentation destroys particles with sizes $a > 1 \mathrm{~mm}$ and redistributes their mass in smaller grains, substructures may appear at millimetre wavelengths outside of the quiescent water snowline (see Fig.~\ref{fig:Fig4_DustPy_SpIndex} and Fig.~\ref{fig:Fig6_SpIndex_Turbulence}). The observation of the fragmentation front could provide key constrains, such as the position of the quiescent snowline or the time passed since the outburst onset \citep[see also][]{houge2023collisional}. More importantly, it could provide a further understanding concerning the fragmentation velocity $v_\mathrm{frag}$ of ice-rich and ice-poor dust particles. In fact, there are uncertainties concerning the fragmentation velocity of water-rich particles, with laboratory experiments finding $v_\mathrm{frag}=10\mathrm{~m~s^{-1}}$ \citep[e.g.][]{supulver1997sticking, gundlach2014stickiness}, while recent works rather find $v_\mathrm{frag}=1\mathrm{~m~s^{-1}}$ similarly to the value associated with dry silicates \citep[e.g.][]{gundlach2018tensile, musiolik2019contacts}. As the fragmentation front is only present if there is a modification of the fragmentation velocity when water ice sublimates, its observation could provide evidence that water-rich particles are indeed more resistant then dry silicates towards fragmentation. Note, however, that it would require high resolution observations, as in our model of V883 Ori the fragmentation front propagates slowly. After $10\,000\mathrm{~yr}$, it is still within $\approx 10 \mathrm{~au}$ for $\alpha_\mathrm{T}=10^{-3}$ (Fig.~\ref{fig:Fig5_Res_DustPy}) and $\approx 30 \mathrm{~au}$ for $\alpha_\mathrm{T}=10^{-2}$ (Fig.~\ref{fig:FigA_Res_MS}). 

Finally, we note that the dark annulus visible in Fig.~\ref{fig:fig1_final_images} around $42 \mathrm{~au}$ does not have a counterpart in the spectral index profiles (see Fig.~\ref{fig:fig3_SpInd_profile}), suggesting that it is not related to a local change in dust particle size or properties. Instead, the dark feature could be caused for example by a dip in the dust temperature or surface density. In the context of the models presented here, such features might be attributed to (a) shadowing of a disc region by the dense inner disc which contains a large population of small grains due to the fragmentation front (see Fig.~\ref{fig:Fig5_Res_DustPy}), or (b) radial drift clearing an annular region of mm-sized particles. Detailed radiative transfer calculations, including evolving dust size distributions and covering the optically thick inner disc, are necessary to test these ideas{, and discard its direct association with the water snowline and/or ongoing planet formation.}

\subsection{Accretion luminosity variability}
\label{sec:discussion_variability}

In this paper, following previous modelling efforts \citep[e.g.][]{schoonenberg2017pebbles, wiebe2019luminosity, vorobyov2022evolution, houge2023collisional}, we assume a step-like function for the temporal evolution of the disc temperature: the temperature instantaneously rises from the quiescent (Eq.~\ref{eq:quiescent_temperature_profile}) to the outburst value (Eq.~\ref{eq:outburst_temperature_profile}), and remains constant throughout the outburst. However, the accretion luminosity and disc temperature during FUor-type accretion outbursts is expected to decay with time passing from the onset \citep[]{audard2014episodic}. In V883 Ori, the current bolometric luminosity is $\sim 218 \mathrm{L_\odot}$ \citep[]{furlan2016herschel}, while it was measured to be $\sim 400 \mathrm{L_\odot}$ about $20$ years ago \citep[]{sandell2001similarity}, and was likely higher at its onset, before 1888 {\citep[]{pickering1890detection, allen1975haro, strom1993discovery}}. In addition to a long-term decay, FU Orionis objects can exhibit low-amplitude variability on short timescales  \citep[$\sim$weeks, see][]{siwak2018insights, takagi2018spectroscopic}, although \citet[]{wendeborn2020study} found that low-amplitude variability in the millimetre emission of FU Orionis objects is rare and, if present, associated with free-free emission. Extending models like the ones presented here to include a more complex luminosity and gas and dust temperature evolution could lead to interesting behavior, and will be the subject of future works. For example, the heating and cooling rates of dust particles are dependant on the particle size \citep[e.g.][]{wolf2003efficient, gavino2023shaping}, so that the onset of ice sublimation may occur at different times depending on the particle size. Moreover, the freeze-out timescale of water molecules can be long (${\sim}100{-}1000 \mathrm{~yr}$), which could lead to a decoupling in between the position at which the dust temperature equals the sublimation temperature of water, and the location where water molecules are actually found dominantly in the ice phase \citep[]{jorgensen2013recent,visser2015chemical, hsieh2019chronology}.

\section{Summary and outlook}
\label{sec:conclusions}

We present new analysis of archival high resolution ALMA observations at $0.88 \mathrm{~mm}$ (Band 7), $1.3 \mathrm{~mm}$ (Band 6), $2.0 \mathrm{~mm}$ (Band 4) and $3.1 \mathrm{~mm}$ (Band 3) of the continuum emission in the protoplanetary disc around V883 Ori, currently undergoing a powerful accretion outburst. From the multi-wavelength data, we obtain new images of the disc (Sect.~\ref{sec:results_final_images}), and constrain the radial profile of the dust spectral index at $0.88 - 1.3 \mathrm{~mm}$, $1.3 - 2.0 \mathrm{~mm}$, and $2.0 - 3.1 \mathrm{~mm}$ (Sect.~\ref{sec:obs_results_spectral_index}). We performed 1D dust evolution simulations, using \texttt{DustPy}, to constrain the state of the dust population in V883 Ori, considering multiple scenarios for the response of dust particles to sublimation of their water ice mantle (Sect.~\ref{sec:results_manyseeds} and Sect.~\ref{sec:results_resilient}). We then qualitatively compared results of the dust evolution models with the observed spectral indices (Sect.~\ref{sec:dustpy_spectralindex}). Our findings are summarized as follows:

\begin{enumerate}[label={\arabic*.}]
    
    \item We obtain new images of the dust thermal emission in V883 Ori in four bands, and observe a dark annulus around $42 \mathrm{~au}$ in Band 4 and Band 7 (Fig~\ref{fig:fig1_final_images}), as was also seen in Band 6 by \citet[]{cieza2016imaging}. 

    \item The inner $20 \mathrm{~au}$ of V883 Ori are characterised by spectral indices $\alpha < 2$, indicative of high optical depths dominated by self-scattering. Outside of $50 \mathrm{~au}$, the emission is optically thin with spectral indices around $\alpha \approx 2.5$ (Fig.~\ref{fig:fig3_SpInd_profile}). The dark annulus visible in Fig.~\ref{fig:fig1_final_images} around $42 \mathrm{~au}$ has no strong imprints in the spectral index profiles.
    
    \item Our relatively low spectral indices suggest the presence of aggregates larger than $1 \mathrm{~cm}$, unlike what was previously assumed in V883 Ori based on Band 6 measurements \citep[$\approx 300\mathrm{~\mu m}$, see][]{cieza2016imaging, schoonenberg2017pebbles}.

    \item Detailed dust evolution models show that, in the sublimation region ($3 < r < 80 \mathrm{~au}$), the evolution of the dust size distribution during an outburst is set primarily by how dust aggregates respond to ice sublimation at the outburst onset. The dust distribution is not in coagulation/fragmentation equilibrium, and particles will either be smaller than $10 \mu$m (many-seeds scenario, Fig.~\ref{fig:Fig4_MS_DustPy}) or larger than $1 \mathrm{~cm}$ (resilient scenario, Fig.~\ref{fig:Fig5_Res_DustPy}).

    \item The resilient scenario is best suited to reproduce the spectral indices observed in V883 Ori ( Fig.~\ref{fig:Fig4_DustPy_SpIndex}), as the many-seeds scenario inevitably leads to spectral indices that are too high (i.e. $\alpha \approx 3.5$), even with additional simulations performed with different turbulence parameters (Fig.~\ref{fig:Fig6_SpIndex_Turbulence}). As large particles were also found in the outbursting disc of FU Ori \citep[]{liu2021millimeter}, we conclude that dust particles survive the sublimation of their water ice mantle at the onset of accretion outbursts (Sect.~\ref{sec:discussion_dust_sublimation}).

\end{enumerate}
Our findings illustrate how constraints on particle size distributions in outbursting systems can be used to study the mechanical and collisional properties of dust aggregates. As FUor-type accretion outbursts are powerful heating events, our findings may have implications for the ice sublimation process in other environments where heating is more gentle. {For example, this could apply to icy pebbles accreting within the envelopes of protoplanets \citep[]{johansen2023anatomy}}, or for icy pebbles drifting through the water snowline in quiescent protoplanetary discs, which may have consequences for the formation pathway of dry planetesimals (Sect.~\ref{sec:discussion_drift_snowline}). Future studies could lead to a more complete picture, for example by focusing on N-body simulations and laboratory experiments to explore how sublimation may vary for different particle sizes, porosities, and water contents, and assessing the plausibility of the reconnection hypothesis (Sect.~\ref{sec:discussion_dust_structure}). Alternatively, higher resolution or longer wavelengths observations contrasting quiescent and outbursting discs may bring key constraints on the fragmentation velocity of ice-rich vs. ice-poor particles (Sect.~\ref{sec:discussion_dustsubstructures}).

\section*{Acknowledgements}

{
We are grateful to the anonymous reviewer for their insightful comments, which helped improve the manuscript. We thank Margot Leemker, Gerhard Wurm, and Jenny Frediani for useful discussions and comments. We also thank Lucas Cieza for sharing their data. This paper highly benefited from the weekly meetings with the Star and Planet Formation group at the European Southern Observatory. This project has made use of the modules \texttt{DSHARP-OPAC} \citep[][]{birnstiel2018disk} and \texttt{DustPy} \citep[][]{stammler2022dustpy}, along with the following Python packages: \texttt{ASTROPY} \citep[][]{robitaille2013astropy}, \texttt{MATPLOTLIB} \citep[][]{hunter2007matplotlib}, and \texttt{NUMPY} \citep[][]{harris2020array}. This paper makes use of the following ALMA data: ADS/JAO.ALMA\#2013.1.00710.S, JAO.ALMA\#2015.1.00350.S, JAO.ALMA\#2015.1.00041.S, JAO.ALMA\#2016.1.00728.S, and JAO.ALMA\#2017.1.00388.S. ALMA is a partnership of ESO (representing its member states), NSF (USA) and NINS (Japan), together with NRC (Canada), MOST and ASIAA (Taiwan), and KASI (Republic of Korea), in cooperation with the Republic of Chile. The Joint ALMA Observatory is operated by ESO, AUI/NRAO and NAOJ.
}

\section*{Data Availability}

The data generated by our dust evolution simulations based on \texttt{DustPy} will be shared upon reasonable request. The ALMA observations used in this manuscript are publicly available on the ALMA archive.


\bibliographystyle{mnras}
\bibliography{paper2} 


\appendix

\section{Dust density distribution for different turbulence parameters}
\label{sec:appendix_1}
\begin{figure*}
    \includegraphics[width=\textwidth]{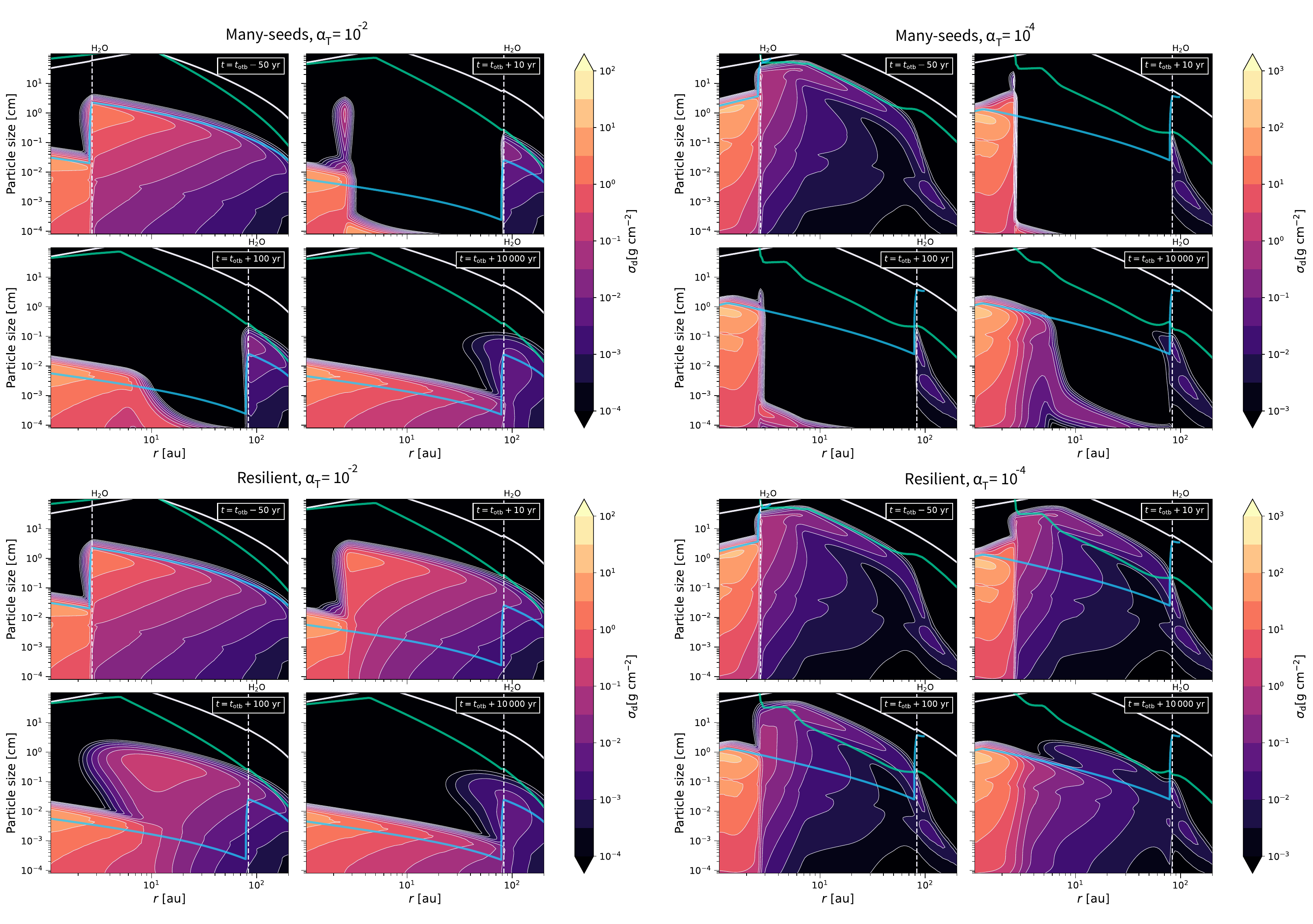}
    \caption{Dust density distribution as a function of particle size in the many-seeds (top) and resilient scenarios (bottom) for $\alpha_\mathrm{T}=10^{-2}$ (left) and $\alpha_\mathrm{T}=10^{-4}$ (right). The colormap refers to the surface density of dust in a given size range at a given radius. In each case, the four panels indicate different key times of the simulation: before the outburst (top-left), after an outburst duration of $10 \mathrm{~yr}$ (top-right), after an outburst duration of $100 \mathrm{~yr}$ (bottom-left), and after an outburst duration of  $10\,000 \mathrm{~yr}$ (bottom-right). The vertical white dotted line ($\mathrm{H_2O}$) specifies the current snowline position. The blue and green solid lines respectively represent the fragmentation and radial drift growth barriers. The solid white line indicates the size at which particles have a Stokes number $\mathrm{St}=1$.}
    \label{fig:FigA_Res_MS}
\end{figure*}

We discuss in this Appendix the evolution of the dust density distribution in the many-seeds {and resilient scenarios for different values of the turbulence, presented in Fig.~\ref{fig:FigA_Res_MS} respectively for} $\alpha_\mathrm{T}=10^{-2}$ and $\alpha_\mathrm{T}=10^{-4}$. Their respective spectral indices at $0.88 - 1.3 \mathrm{~mm}$ are presented in Fig.~\ref{fig:Fig6_SpIndex_Turbulence}. When comparing the pre-outburst distributions {for the different turbulence strength (left vs. right columns in} Fig.~\ref{fig:FigA_Res_MS}), we notice how the turbulence impacts the evolution of the dust population. Stronger turbulence ($\alpha_\mathrm{T} = 10^{-2}$) leads to higher relative velocities between particles, which shifts the onset of fragmentation to smaller sizes $a_\mathrm{max}$ \citep[]{birnstiel2011dust}. The radial transport is reduced, and after almost $t_\mathrm{otb}=10^{5} \mathrm{~yr}$, a significant amount of dust mass persists above $r>10 \mathrm{~au}$. The opposite is visible for lower turbulence ($\alpha_\mathrm{T} = 10^{-4}$), where pebbles drift efficiently, which halts the redistribution of mass in smaller fragments and decreases the surface density in mm-sized particles.

When the outburst starts, the turbulence mostly influences the dust evolution due to the modified collision rates. The outward propagation of the growth and fragmentation fronts detailed in Sect.~\ref{sec:results_manyseeds} and Sect.~\ref{sec:results_resilient} is faster for larger turbulence. In fact, after an outburst duration of $10\,000 \mathrm{~yr}$ {(bottom-right panels in the left column in} Fig.~\ref{fig:FigA_Res_MS}), it is above $30 \mathrm{~au}$. At this stage, the density distribution is almost identical in both scenarios and settle in the new coagulation/fragmentation equilibrium, so that it is no longer possible to disentangle how icy aggregates responded to sublimation. However, such a long duration is unexpected for an FUor-type accretion outburst, so that in outbursting discs we would always expect the many-seeds and resilient scenarios to show different density distributions (see Sect.~\ref{sec:dustpy_results}).

\bsp    
\label{lastpage}
\end{document}